\documentclass[fleqn,usenatbib, onecolumn]{mnras}

\usepackage{newtxtext,newtxmath}
\usepackage{multirow}

\usepackage[T1]{fontenc}

\DeclareRobustCommand{\VAN}[3]{#2}
\let\VANthebibliography\thebibliography
\def\thebibliography{\DeclareRobustCommand{\VAN}[3]{##3}\VANthebibliography}

\usepackage{graphicx}	
\usepackage{amsmath}	
\usepackage{xcolor}
\usepackage{soul}
\usepackage{booktabs}
\usepackage{bm}

\definecolor{forestgreen}{rgb}{0.13, 0.545, 0.13}

\newcommand{\sky}{{\texttt{skyFACT}}}

\title[Morphology of the GCE]{Robust inference of the Galactic centre gamma-ray excess spatial properties}

\author[D.~Song et al.]  
{Deheng Song$^{1}$\thanks{E-mail: songdeheng@yukawa.kyoto-u.ac.jp},
Christopher Eckner$^{2,3,4}$,
Chris Gordon$^{5}$,
Francesca Calore$^{3}$,
Oscar Macias$^{6,7}$,
\newauthor
Kevork N.\ Abazajian$^{8}$,
Shunsaku Horiuchi$^{9,10}$,
Manoj Kaplinghat$^{8}$,
Martin Pohl$^{11,12}$
\\
$^{1}$Center for Gravitational Physics and Quantum Information,
Yukawa Institute for Theoretical Physics, Kyoto University, Kyoto 606-8502, Japan\\
$^{2}$Center for Astrophysics and Cosmology, University of Nova Gorica, Vipavska 11c, 5270 Ajdov\v{s}\v{c}ina,
Slovenia\\
$^{3}$LAPTh, CNRS,  USMB, F-74940 Annecy, France\\
$^{4}$LAPP, CNRS,  USMB, F-74940 Annecy, France\\
$^{5}$School of Physical and Chemical Sciences, University of Canterbury, Christchurch, New Zealand\\
$^6$Department of Physics and Astronomy, San Francisco State University, San Francisco, California 94132, USA\\
$^7$GRAPPA Institute, University of Amsterdam, 1098 XH Amsterdam, The Netherlands\\
$^8$Department of Physics and Astronomy,
University of California - Irvine, Irvine, California 92697, USA\\
$^9$Center for Neutrino Physics, Department of Physics, Virginia Tech, Blacksburg, VA 24061, USA \\
$^{10}$Kavli Institute for the Physics and Mathematics of the Universe (WPI), University of Tokyo, Chiba 277-8583, Japan\\
$^{11}$Deutsches Elektronen-Synchrotron DESY, Platanenallee 6, 15738 Zeuthen, Germany\\
$^{12}$Institute of Physics and Astronomy, University of Potsdam, 14476 Potsdam, Germany
}

\date{}
\begin{document}
\label{firstpage}
\pagerange{\pageref{firstpage}--\pageref{lastpage}}
\maketitle

\begin{abstract}
The gamma-ray {\it Fermi}-LAT Galactic centre excess (GCE) has puzzled scientists for over 15 years. Despite ongoing debates about its properties, and especially its spatial distribution, its nature remains elusive.
We scrutinize how the estimated spatial morphology of this excess depends on models for the Galactic diffuse emission, focusing particularly on the extent to which the Galactic plane and point sources are masked.
Our main aim is to compare a spherically symmetric morphology—potentially arising from the annihilation of dark matter (DM) particles—with a boxy morphology—expected if faint unresolved sources in the Galactic bulge dominate the excess emission.
Recent claims favouring a DM-motivated template for the GCE are shown to rely on a specific Galactic bulge template, which performs worse than other templates for the Galactic bulge.
We find that a non-parametric model of the Galactic bulge derived from the VVV survey results in a significantly better fit for the GCE than DM-motivated templates.
This result is independent of whether a \texttt{GALPROP}-based model or a more non-parametric ring-based model is used to describe the diffuse Galactic emission. This conclusion remains true even when additional freedom is added in the background models, allowing for non-parametric modulation of the model components and substantially improving  the fit quality. When adopted, \emph{optimized} background models provide robust results in terms of preference for a boxy bulge morphology for the GCE, regardless of the mask applied to the Galactic plane.  
\end{abstract}




\section{Introduction}

The successful deployment of the {\it Fermi} Gamma-Ray Space Telescope fifteen years ago ushered in an era of unprecedented sensitivity to the gamma-ray sky, with {\it Fermi}'s Large Area Telescope (LAT) providing increased energy resolution and an unprecedented 
 angular resolution	
\citep{Fermi-LAT:2009ihh}. A key science goal of \textit{Fermi}-LAT was to explore gamma-ray emissions from dark matter (DM), specifically investigating the pair annihilation products of thermally-produced weakly-interacting massive particles (WIMPs).

Shortly after its launch, an observation was reported of an extended source toward the Galactic centre (GC) consistent with thermal WIMP annihilation with a profile consistent with a cuspy DM halo \citep{Goodenough:2009gk,Vitale2009,Hooper:2010mq}.
The presence of excess emission was confirmed and refined with many subsequent studies
\citep[e.g., ][]{Hooper:2011ti, Abazajian:2012pn, Hooper:2013rwa, Gordon:2013vta, Abazajian:2014fta, Daylan:2014rsa, Calore:2014xka, Zhou:2014lva, TheFermi-LAT:2015kwa, Linden:2016rcf,Fermi-LAT:2017opo}.

This GC excess (GCE) may also be due to an unresolved population of, e.g., millisecond pulsars (MSPs), stellar remnants associated with the central stellar population of the Milky Way \citep{Abazajian:2010zy, Abazajian:2012pn}.
However, there is debate on whether more GC MSPs should have been resolved \citep[e.g., ][]{hooperGammarayLuminosityFunction2016,ploegComparingGalacticBulge2020, Dinsmore:2021nip, Malyshev24} and whether there are sufficient corresponding low mass X-ray binary (LMXB) systems \citep[e.g.,][]{Haggard2017,gautamMillisecondPulsarsAccretion2022}.

One method of distinguishing between the MSP and DM proposals for the GCE is whether the GCE has a spherical morphology \citep{Daylan:2014rsa, Calore:2014xka, TheFermi-LAT:2015kwa, DiMauro2021, Cholis2022,McDermott2023} or follows the bulge-like morphology of the GC's stellar populations \citep{Macias:2016nev, Bartels:2017vsx, Macias:2019omb, Abazajian:2020tww, Calore:2021jvg, 2022ApJ...929..136P}. The main obstacle to accurately determining the GCE morphology is uncertainties in the Galactic diffuse emission. 
When residuals in the gamma-ray fit were significantly reduced either through the construction of advanced gas models \citep{Macias:2016nev, Macias:2019omb, 2022ApJ...929..136P} or through new, more flexible, fitting techniques~\citep{ Bartels:2017vsx,Calore:2021jvg}, a strong preference for the excess tracing old stars in the Galactic bulge emerged.
Recently, \cite{DiMauro2021,Cholis2022,McDermott2023} have, however, made opposite claims.

Another, independent, way of discriminating between the DM and MSP explanations of the GCE is to look for non-Poissonian statistics in the GCE, which would be indicative of the MSP explanation \citep{Bartles2016,Lee:2015fea}, in constrast to any other truly-diffuse signal such as DM. However, this is arguably even more susceptible to uncertainties in the Galactic diffuse emission \citep{Leane:2019xiy, Chang:2019ars, Buschmann:2020adf, Leane:2020pfc, Leane:2020nmi}. 
New and independent analyses, introducing methodological developments which effectively reduce the impact of Galactic diffuse emission modelling systematics, found a sizable contribution from faint, sub-threshold point sources to the GCE 
~\citep{Calore:2021jvg,List:2020mzd,List:2021aer,Mishra-Sharma:2021oxe}.

Accounting for uncertainties in the Galactic diffuse emission model is undoubtedly key to making robust inferences on the Fermi GCE properties and overcoming the so-called ``reality gap'', i.e. the discrepancy between models and real data~\citep{Caron:2022akb}.
Different solutions were explored in the literature to overcome systematics related to the Galactic diffuse emission modelling. These approaches either rely on the input from a large number of realisations 
of cosmic-ray induced diffuse gamma-ray emission as obtained from cosmic-ray propagation codes, like \texttt{GALPROP}\footnote{http://galprop.stanford.edu} \citep{Strong:1998pw}, \citep[e.g., ][]{Calore:2014xka, Cholis2022}, or they allow more flexibility in the diffuse-emission model by splitting the Galactic-diffuse-emission templates in multiple rings \citep[e.g., ][]{Macias:2016nev, DiMauro2021}.
Finally, a complementary way to optimize gamma-ray emission model components and, more specifically, the Galactic diffuse emission is the application of data-driven techniques that can reduce the residuals and minimize the gap between model space and reality. A tool that has been utilized in the context of the GCE is \texttt{skyFACT} \citep{Storm:2017arh, Bartels:2017vsx, Calore:2021jvg}. \texttt{skyFACT} combines the capabilities of traditional template-based maximum likelihood fits with image reconstruction techniques. Its advantage lies in the addition of spatial and spectral re-modulation for all components of the compiled gamma-ray emission model. In this sense, \texttt{skyFACT} allows for a more flexible treatment of the employed templates, especially regarding their spatial morphology, which is fixed in standard template-based fits. To achieve this flexibility, \texttt{skyFACT} introduces a large number of nuisance parameters and a penalizing likelihood function constraining their variation during the fit in addition to the Poisson likelihood function typically adopted in gamma-ray analyses; for technical details, see \cite{Storm:2017arh}.

With this work, we aim at systematically scrutinising how the morphology of the GCE is affected by background modelling uncertainties when point sources and the disk plane are masked, and analysis approaches are varied, and address in detail some contradictory findings in the recent literature, most notably those in~\cite{McDermott2023} (M2023 hereafter). As such, our analysis is solely focused on the large-scale emission properties of the GCE and not on its point-source contribution.

In Section~\ref{sec:models}, we explain the model components and fitting procedure, exploring the background model construction.
In Section~\ref{sec:M2023}, we focus on template fitting.  We show that the ring-based fits of M2023 have not properly converged. We also provide evidence that they have used a bulge template that is inconsistent with the data. 
We show that by switching to a more accurate bulge template, such as the one generated from the recent VISTA Variables in the Via Lactea (VVV) survey \citep{Coleman:2019kax}, the bulge is favoured over DM-based templates even when M2023's background is used.
In Section~\ref{sec:Masked Analysis Using Data-Driven Diffuse Galactic Emission  Templates}, we show that the ring-based methods still find a preference for the Galactic bulge over the DM template even when the point sources masks are substantially increased in size. We also show that after masking, the inclusion of a Galactic bulge passes a Monte Carlo based goodness of fit test. In Section~\ref{sec:skyfact_analysis}, we employ a  \sky~modulation 
of the diffuse templates \citep{Storm:2017arh, Bartels:2017vsx, Calore:2021jvg}. We find that in all cases, the Bayesian evidence is improved by the modulation, and the addition of the DM template is not favoured once the modulation has been done and a Galactic bulge template has been included. The conclusions are given in Section~\ref{sec:conclusions}.

\section{Gamma-ray model components and fitting procedure}
\label{sec:models}
In this section, we first provide a brief overview of the components used to model the gamma-ray sky.
In general, the inner Galaxy sky is interpreted as the sum of the following main contributions:
\begin{itemize}
    \item[(i)] Galactic diffuse emission: This contribution, which we will discuss in more detail below, is the result of the interactions of cosmic rays with the interstellar gas and low-energy interstellar radiation fields. 
    The main contributors to the Galactic diffuse emission are the decay of neutral pions produced in collisions between cosmic-ray protons and interstellar gas, the inverse Compton scattering of the interstellar radiation field by electrons, and the bremsstrahlung emission from these electrons.
    \item[(ii)] Point-like and extended sources: These correspond to gamma-ray identified sources which are listed in the {\it Fermi}-LAT catalogues and can be masked or refitted in the analysis. We will discuss the treatment of point-like sources in the following sections. 
    \item[(iii)] Isotropic diffuse gamma-ray background emission (IGRB): This component accounts for the (almost) isotropic emission measured at high latitudes and is thought to originate from the superposition of different contributions mostly from extragalactic, faint sources~\citep{Fermi-LAT:2014ryh}.
    \item[(iv)] Galactic centre excess, GCE: To complete the description of the inner Galaxy gamma rays, it has been demonstrated that one needs to consider an additional contribution, the so-called GCE. 
\end{itemize}

\subsection{The Galactic diffuse emission models}
Given the vast number of independent degrees of freedom across the sky, constructing an expected Galactic diffuse emission model requires numerous assumptions. Works like \cite{Cholis2022} (hereafter C2022) postulate that the inner galaxy is predominantly influenced by Galactic diffuse emission stemming from nearly steady-state astrophysical processes and modelled through standard propagation codes like \texttt{GALPROP}. In the following we will refer to these models as ``\texttt{GALPROP}-based'' templates. In contrast, other approaches, such as those in \cite{Macias:2016nev, Macias:2019omb, Abazajian:2020tww, 2022ApJ...929..136P}, introduce concentric cylindrical, galactocentric-based templates, which we will refer to as ``ring-based templates''.

There are conflicting reports in the literature regarding the best method to use for modeling the Galactic diffuse emission. First, on the background model construction, \cite{2022ApJ...929..136P} (hereafter P2022\footnote{Note that there is some overlap in the authors of the current article and those of P2022.}) demonstrated that the ring-based method provided a better fit to the \textit{Fermi}-LAT data than the non-ring-based method of \cite{DiMauro2021}. This was followed by M2023, who reported that when the point sources and Galactic plane were masked, the \texttt{GALPROP}-based templates, generated by C2022, provided a better fit to the \textit{Fermi}-LAT data. Second, there are conflicting reports of whether the GCE is spherical DM-like or bulge-like. On one hand, P2022 found that bulge models were better explanations of the data, while, on the other hand, M2023 claimed a better fit for the spherical DM template as opposed to the Galactic bulge. 

In this work, we systematically compare ``\texttt{GALPROP}-based'' and ``ring-based'' templates for the Galactic diffuse emission, with the final goal of better understanding some contradictory claims in the literature. 

In summary, the templates that compose the Galactic diffuse emission in the two approaches are:
\begin{description}
\item[{\bf ``\texttt{GALPROP}-based'' model:}] 
C2022 and M2023 made use of \texttt{GALPROP} to propagate cosmic rays within the interstellar medium and evaluated the diffuse emission templates. 
The diffuse emissions of pion decay is combined with the sub-dominant bremsstrahlung contribution into one template. Inverse-Compton (ICS) emission was also included. The templates from \texttt{GALPROP} were augmented by templates for the isotropic background and Fermi bubbles. In total, the ``\texttt{GALPROP}-based'' model has four templates. 
\item[{\bf ``Ring-based'' model}:] 
As in P2022 and M2023, the ring-based approach utilizes 16 independent galactocentric cylinders. This model comprises four rings for the neutral atomic hydrogen (HI) density and another four for the molecular hydrogen (H2) density. Additionally, six rings follow the ICS emission. Also included are two dust-derived templates, one negative and one positive valued. The positive dust correction template physically represents HI and H2 hydrogen that is not traced by the relevant emission, known as the dark neutral medium, or an overestimation of the atomic hydrogen spin temperature \citep{Fermi-LAT:2016zaq}. The negative dust correction template represents an underestimation of the spin temperature. The HI and H2 rings, shaped as annular cylinders, have boundaries located at 3.5, 8, 10, and 50 kpc from the GC. The ICS rings share these boundaries, except the innermost ring is further subdivided at radii of 1.5, 2.5, and 3.5 kpc. Also incorporated are the identical isotropic background and Fermi bubbles template as was done by M2023. A total of 18 templates are used for the ``ring-based'' model.
\end{description}

\subsection{The Galactic centre excess templates}
As done by the majority of the literature, in this work we will consider two main hypotheses for the morphology of the GCE. 

The first one is that the signal is connected with annihilating DM in the inner Galaxy.
To this end, we consider DM-inspired templates constructed as a spherical template that follows the square of a generalized Navarro-Frenk-White profile (gNFW$^2$):
\begin{equation}
\rho(r) \propto \frac{1}{\left(r / r_c\right)^\gamma\left(1+r / r_c\right)^{3-\gamma}}.
\end{equation}
We adopt the parameters used in C2022 and M2023, which are $\gamma = 1.2$ and $r_c = 20$ kpc.
The DM template is then obtained by integrating $\rho(r)^2$ along each line of sight. Note, however, that there is currently no clear prediction for the inner  (within about 2 kpc) density profile of the DM halo of the MW (or its sphericity) with both cusps and cores found in sophisticated hydrodynamic simulations~\citep{Lazar:2020pjs,Grand:2022olu}.

The second hypothesis builds on the fact that the GCE may trace the distribution of old stars in the 
Galactic bulge. 
Direct observations of the GC region have historically been obscured due to challenges posed by dust reddening. The advent of near-infrared surveys, such as the groundbreaking COBE/DIRBE study, has enabled significant advancements in our understanding of this central region \citep{Bland-Hawthorn2016}. These surveys have not only confirmed the presence of the Galactic bulge/bar but also provided the foundational data upon which subsequent triaxial bar models of our galaxy were constructed \citep{BinneyUnderstandingkinematicsGalactic1991,WeilandCOBEDiffuseInfrared1994,Freudenreich:1997bx}. \cite{2013MNRAS.434..595C}  utilized red clump giants from the OGLE-III survey to create a detailed photometric model of the Galactic bar. This relied on data available up to 2013. In an advancement, \cite{Coleman:2019kax} utilized more contemporary data from the VVV survey. Their approach, integrating non-parametric methodologies, offers a 
flexible means of estimating the morphology of the Galactic bulge, thus refining our understanding of this critical Galactic structure.
The flexibility is important because we want the data and not an assumed inflexible functional form, as in  \cite{2013MNRAS.434..595C} for example, to determine the radial profile of the template.

\begin{figure}
	\centering
	\includegraphics[width=0.32\columnwidth]{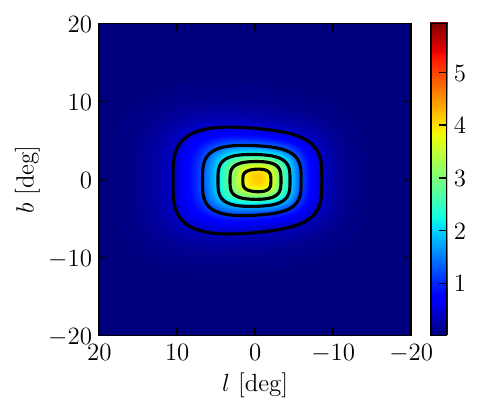}
	\includegraphics[width=0.32\columnwidth]{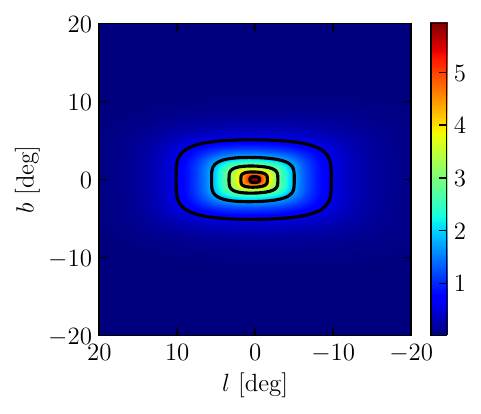}
	\includegraphics[width=0.32\columnwidth]{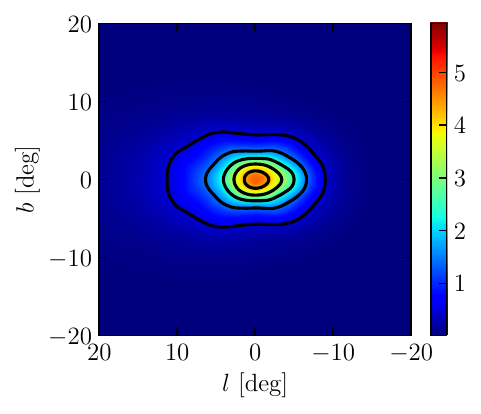}
	\caption{Spatial templates of the Galactic bulge models considered in this work. References from left to right: ~\citet{Freudenreich:1997bx},~\citet{2013MNRAS.434..595C}, and~\citet{Coleman:2019kax}.
		Note that these are line-of-sight integrated images of the bulge templates before they are convolved with the PSF.
	} 
	\label{fig:bulge_templates}
\end{figure}

In the present work, we consider several different bulge templates. 
The bulge model used in M2023 is publicly available as part of the analysis package \texttt{gcepy}\footnote{\url{https://github.com/samueldmcdermott/gcepy}}. Following M2023, we will label this the boxy-bulge BB (\texttt{gcepy}) template. 
In addition to one from \texttt{gcepy}, we consider three other bulge models:  \citet{Freudenreich:1997bx} (hereinafter F98), \citet{2013MNRAS.434..595C} (hereinafter Cao13), and \citet{Coleman:2019kax} (hereinafter Coleman20). 
We notice that the BB template in \texttt{gcepy} appear be obtained from the Cao13 model (M2023, personal communication),
but upon further inspection, we found that it does not match our version of the Cao13 template. We therefore keep the BB \texttt{gcepy} template and the Cao13 as different, independent, choices for the bulge. 
Figure~\ref{fig:bulge_templates} shows the spatial templates for these bulge models in the main region of interest (ROI) of this work, i.e., 40$^\circ$ $\times$ 40$^\circ$ around the GC. We have normalized the templates to have the same flux in the ROI (in an arbitrary unit). The contours on the maps show the 10\%, 30\%, 50\%, 70\%, and 90\% levels with respect to the central value. The contours demonstrate the differences in the bulge models, both at large scales as well as near the centre. 

On top of the boxy-bulge, we also add a component for the nuclear bulge (NB) following the parametric model of \citet{2002AA...384..112L}, which is
necessary to include when not masking the Galactic plane, see Section~\ref{sec:Masked Analysis Using Data-Driven Diffuse Galactic Emission  Templates} and Section~\ref{sec:skyfact_analysis}.

\subsection{Fitting procedure and statistical framework}
\label{sec:fitting-and-statistics}

In this paper, we adopt two different fitting techniques. 
The first one is the traditional ``template fit'', where all model components are defined as morphological templates that are not allowed to vary during the fit procedure.
In template fitting, the free parameters are the independent bin-by-bin normalizations of the spectral energy 
distribution of each model component. The second technique is the so-called adaptive template fitting, where the spatial distribution of photons among different sky components is enabled to be re-modulated during the fitting procedure~\citep{Storm:2017arh, Bartels:2017vsx, Calore:2021jvg}. In what follows, we briefly introduce more technical details about the implementation of both data fitting approaches. We conclude this section by outlining the statistical inference methods that we apply to investigate the morphology of the GCE.

\subsubsection{Traditional template fitting}
\label{sec:template-fit-technique}
In standard template fit analysis, the gamma-ray sky is described as the sum of multiple model components, $k$,
identified by a fixed spatial distribution, i.e.~the spatial template. 
The model $\phi$ is a linear combination of $k$ spatial templates $T$ binned in spatial pixels $p$ and a spectral normalization $\Sigma$ for each energy bin $b$, which is free to re-adjust during the fit: 
\begin{equation}
    \phi_{pb} = \sum_{k} T^{(k)}_{p}\cdot \Sigma_{b}^{(k)} \mathrm{.}
\end{equation}
The optimization of the free energy-independent normalization is done by maximizing the Poissonian likelihood given the number count maps. 

For the optimization, M2023 employs \texttt{dynesty} to scan the parameter space and then uses a No-U-Turn sampler (utilizing \texttt{numpyro}) to optimize the best-fit result.
For the sake of cross-checking the M2023 results, in Section~\ref{sec:M2023}, we will perform the maximum likelihood analysis with an alternative minimizer, i.e.~using the Limited memory Broyden–Fletcher–Goldfarb–Shanno algorithm extended to handle simple box constraints (L-BFGS-B). For implementation, we use the Python package \texttt{lmfit}. We use \texttt{stats.poisson.logpmf} provided by \texttt{scipy} to calculate the Poissonian log-likelihood. Following M2023, 
 we also include penalty terms (provided by \texttt{gcepy}) on the log-likelihood to account for  when the IGRB and Fermi bubbles normalizations deviate too much from their spectra measured at high latitudes. 

Point sources in the 4FGL are masked for this analysis. We will provide more details on the mask implementation in Section~\ref{sec:M2023}.

\subsubsection{Adaptive-constrained template fitting}
\label{sec:skyfact-technique}

A complementary analysis technique to investigate the preferred morphology of the GCE is adaptive template fitting implemented in the software package \texttt{skyFACT}. As mentioned in the introduction, \texttt{skyFACT} is a mixture of a conventional template-based fit and more advanced image reconstruction routines. The main advantage is a departure from imposing a non-flexible prior on the spatial morphology of the compiled background and signal components of the gamma-ray emission model. While this flexibility enables capturing and remedying a certain degree of component mis-modelling, it requires the introduction of a large number of nuisance parameters that have to be controlled during the fit to avoid over-fitting and unphysical results. Therefore, \texttt{skyFCAT} utilizes a combination of a Poisson and a penalizing likelihood function to guide the fit with constrained freedom for all nuisance parameters. The fit itself is the minimization of the mentioned log-likelihood function.

In essence, the gamma-ray emission model compiled for \texttt{skyFACT} is identical to the input data required for a traditional template fit. That is, the model $\phi$ is a (tri-)linear combination of $k$ spatial templates $T$ binned in spatial pixels $p$ and a spectral normalization $S$ per energy bin $b$ following a certain functional form or tabulated data:
\begin{equation}
    \phi_{pb} = \sum_{k} \tau_{p}^{(k)}T^{(k)}_{p}\cdot\sigma_{b}^{(k)}S^{(k)}_{b} \cdot \nu^{(k)}\mathrm{.}
\end{equation}
Examples of suitable spatial templates are the Galactic bulge models in Figure~\ref{fig:bulge_templates}. The addition of \texttt{skyFACT} is to introduce a global normalization parameter $\nu$ per component and spatial and spectral nuisance parameters, $\tau$ and $\sigma$, respectively. In practice, $\tau$ and $\sigma$ act as a one-to-one copy in shape of the spatial and spectral input priors. Both nuisance parameter sets are initialized with value 1 and varied -- or better, \textit{re-modulated} -- in the priors' stead during the fitting procedure. The nuisance parameters are required to be strictly positive. 

We note that the \texttt{skyFACT} model $\phi$ allows for the inclusion of point-like sources according to their position and spectra listed in gamma-ray catalogs of choice. For them, the position inside the considered region of interest is fixed, i.e.~there is no associated spatial template $T$. Consequently, all injected point-like sources are spectrally re-fit in adaptive template fitting.

The numerous nuisance parameters are tamed via the regularizing part of \texttt{skyFACT}'s likelihood function, which depends on five hyper-parameters per component\footnote{Only three in the case of point-like sources.} constraining the magnitude of individual parameters and the correlation among neighbouring parameters. All details about the explicit structure of the likelihood function and hyper-parameters is provided in \cite{Storm:2017arh}. Here, it shall suffice to give an illustrative example of the functionality of two of them, the spatial and spectral smoothing scale $\eta$ and $\lambda$. They are defined as $\eta = 1/x^2$ ($\lambda = 1/E^2$), where $x$ ($E$) is the allowed variation between neighbouring spatial pixels (energy bins). Setting the smoothing scales to zero is thus equivalent to saying that all spatial pixels (energy bins) may vary entirely independently of each other. 

As the tuning of the hyper-parameters is cumbersome due to computational speed, we select the configuration of \texttt{run5} in \cite{Storm:2017arh} as the foundation for our settings. The specifications of \texttt{run5} assume a certain spatial smoothing for the brightest gamma-ray components, namely: The inverse Compton component receives the largest smoothing, followed by the $\pi^0$-component and lastly the Fermi Bubbles with the shortest smoothing scale. Regarding the remaining model elements we do not allow for any spatial re-modulation of the input template, this applies to the different GCE components, in particular. We hence ensure that the minimization problem remains convex guaranteeing convergence of the L-BFGS-B algorithm utilized in \texttt{skyFACT}. Spectral smoothing is not applied at all.

\subsubsection{Statistical framework: Parameter inference and model comparison}
\label{sec:statistical-framework}
In a frequentist setting it is possible to properly quantify the preference for different gamma-ray emission models based on the maximum likelihood method and the resulting likelihood values at the optimal point. However, a well-defined notion of model comparison from a statistical point of view is limited to nested models. By nested models we mean any gamma-ray emission model that is comprised of a default set of components (or base components) and to which further components are added without removing any contributions from the default set. Then it is possible to compute the preference in the data for the enlarged model over the base model in terms of significance.  

To be more quantitative let us assume that we add a single component $X$ to the base model and perform a standard template fit so that the extended model has $N$ additional parameters (normalizations per energy bin). Let $\ln{\mathcal{L}_{\mathrm{base}}}$ denote the log-likelihood value evaluated for the best-fitting parameters determined via the maximum likelihood approach. Likewise, $\ln{\mathcal{L}_{\mathrm{base+X}}}$ is the corresponding log-likelihood value found for the extended base model including $X$. We choose the log-likelihood ratio test statistic as a means to quantify the significance: $\mathrm{TS} = 2(\ln{\mathcal{L}_{\mathrm{base+X}}} - \ln{\mathcal{L}_{\mathrm{base}}})$.  Within this setup, the test statistic is distributed according to a mixture distribution following \citep{Macias:2016nev}
\begin{equation}
    p(\mathrm{TS}) = 2^{-N}\left[\delta(TS) + \sum_{k = 1}^{N} \binom{N_{\mathrm{dof}}}{k}\chi^2_{k}(\mathrm{TS})\right]\rm{.}
\end{equation}
Here, $\delta$ refers to the Dirac distribution, $\binom{n}{k}$ is the binomial coefficient and $\chi^2_{k}$ denotes a $\chi^2$-distribution with $k$ degrees of freedom. The significance $\sigma$ of the added component under the observation of the test statistic value $\hat{\mathrm{TS}}$ amounts to
\begin{equation}
\label{eq:significance-mixture-model}
    \sigma(\hat{\mathrm{TS}}) = \sqrt{\mathrm{CDF}^{-1}\!\left(\chi^2_1, \mathrm{CDF}\!\left(p(\mathrm{TS}), \hat{\mathrm{TS}}\right)\right)}\rm{,}
\end{equation}
where $\mathrm{CDF}(f, x)$ refers to the cumulative distribution function of $f$ at $x$ and CDF$^{-1}$ is its inverse.

Model comparison is challenging to correctly perform in a frequentist framework as shown above, while it is better defined and easier to access
in a Bayesian approach.
We also adopt the latter to run model comparison.
Deriving the Bayesian evidence of each individual gamma-ray emission model allows us to perform a Bayesian model comparison or hypothesis testing based on the Bayes factor. Given the Bayesian evidence $\ln{\mathcal{H}_X}$ of model X and $\ln{\mathcal{H}_Y}$ of model Y, the mutual Bayes factor is given by $\mathcal{B}_{XY} = \exp{(\ln{\mathcal{H}_X} - \ln{\mathcal{H}_Y})}$. A positive value of the Bayes factor implies a certain degree of evidence for model X being preferred over model Y. In what follows, we utilize the empirical classification of the degree of evidence from table 1 of \cite{Trotta:2008qt} based on the logarithmic Bayes factor. \footnote{We also provide the definition of TS between different models, as $TS = 2(\ln \mathcal{L}_{X} - \ln \mathcal{L}_{Y})$. While we refrain from assessing model comparison through this $TS$, we notice however that this practice is largely present in the previous literature. Therefore, for the sake of comparison, in the main results' tables we will also report this value without over-interpreting it.}

While the Bayesian framework offers a direct way of stating the preference for one gamma-ray emission model over another one -- irrespective of the exact composition -- it depends by construction on prior probabilities for all parameters. In this sense, it carries a certain intrinsic user bias due to the choice of priors, may it be their parametric shape or range. Consequently, we remark as a caveat that the derived preference for a model is prior-dependent. As we will show later in Section~\ref{sec:M2023-ring}, a suitable choice for the prior range is essential to cover the correct best-fitting point in the model's parameter space. 

\section{Template fitting: Reproducibility of previous works and improvements}
\label{sec:M2023}
In this section, we weigh in on the findings of M2023, first reproducing and then improving on their analysis.  
The fitting technique here adopted is the traditional template fit (Section~\ref{sec:fitting-and-statistics}).
We will perform Bayesian model comparison to assess what is the best model for describing gamma-ray data among the ones we test.
We will also discuss how the evidence for the DM-inspired template is affected by different choices of point sources and Galactic plane mask, by making use of nested models in the frequentist approach.

\subsection{Reproducing M2023}
As a first step, we repeat the analysis performed in M2023, by adopting the same dataset and models.
We remind the reader that the M2023 analysis was performed using both \texttt{GALPROP}- and ring-based background models with standard template fitting, and the authors claimed that the GCE is better described by a DM-like model than a bulge one.
A summary of the data selection in C2022 and M2023 is reported in Table~\ref{tab:data_cholis}.
We adopt this dataset as publicly available through the \texttt{gcepy} webpage. 
Together with the selected data set (counts map), the \texttt{gcepy} package  released also (i) Galactic plane and point-source mask adopted, and (ii) the model templates convolved with the \textit{Fermi}-LAT point spread function (PSF). 
The mask adopted by M2023 (and in this section unless otherwise specified) masks out both the Galactic plane ($|b| < 2^\circ$) and the point sources in the 4FGL-DR2 catalogue. 

In M2023, two \texttt{GALPROP}-based models are identified as best-fit ones: \texttt{GALPROP}$_{7p}$ is, 
according to M2023, the best Galactic diffuse model when no GCE is added to the fit; \texttt{GALPROP}$_{8t}$
instead provides the best performance when adding a GCE modeled through a gNFW$^2$ profile. 
In  C2022's Zenodo archive~\citep{cholis_2022_6423495}, these two cases can be seen to correspond to the cases XLIII and XLIX, respectively, in Table VIII of C2022. 

We run a traditional template fit though maximum likelihood optimization, as described in Section~\ref{sec:fitting-and-statistics}. The minimization is run with both the \texttt{gcepy} code and our implementation with the L-BFGS-B algorithm.

The results from the runs reproducing M2023 are reported in Table~\ref{tab:llh_GALPROP}. 
Following M2023, we use the background model \texttt{GALPROP}$_{8t}$, corresponding to the best-fit model obtained when an additional gNFW$^2$ component has been added. 
As done by M2023, we compare this to the best-fit background in the case of no GCE source (\texttt{GALPROP}$_{7p}$). Compared with the 
scenario without GCE, the BB \texttt{gcepy} template has a $\ln \mathcal{B} = 885$, indicating evidence for 
a GCE. We remind the reader that no NB is included in the model, following M2023.

For \texttt{GALPROP}-based background models we find very similar likelihood values no matter what is the adopted minimization procedure. This contrasts to what happens for ring-based templates (see below). 

\begin{table}
    \centering
        \caption{Data selection according to~C2022 and~M2023.}
    \begin{tabular}{c|c}
    \toprule
    Parameter & Value \\\midrule
    Time range & Week 9 to Week 670 (12.5 years)\\
    Energy range & 14 bins from 0.275 to 51.9 GeV \\
    ROI & 40$^\circ$ $\times$ 40$^\circ$, bin size = 0.1$^\circ$\\
    Data class & \texttt{P8R3\_CLEAN\_V3} (evclass=256), \texttt{FRONT} (evtype=1)\\
    Filter & \texttt{(DATA\_QUAL==1)\&\&(LAT\_CONFIG==1)}\\
     & \texttt{\&\&(ABS(ROCK\_ANGLE)<52)}\\
    Max zenith angle & 100 deg\\
    \bottomrule
    \end{tabular}
    \label{tab:data_cholis}
\end{table}

\subsection{Model systematics I: Testing other bulge templates}
\label{sec:M2023-bulge}
\begin{figure}
	\centering
	\includegraphics[width=0.24\columnwidth]{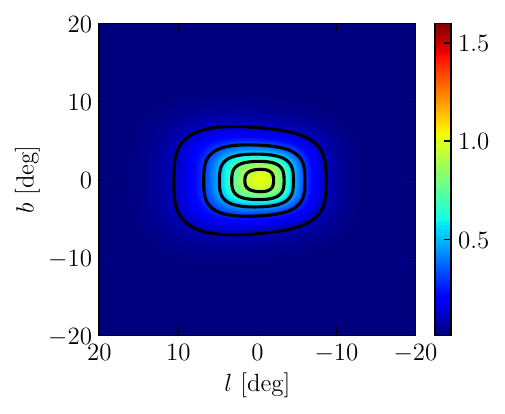}
	\includegraphics[width=0.24\columnwidth]{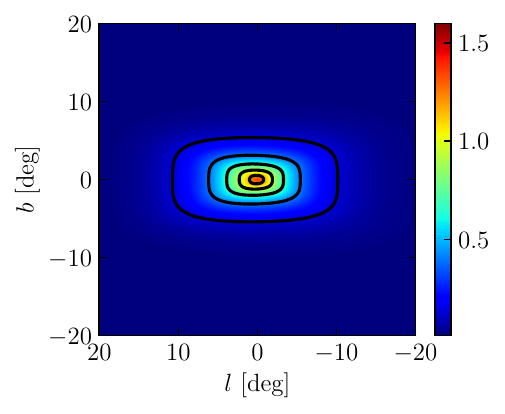}
	\includegraphics[width=0.24\columnwidth]{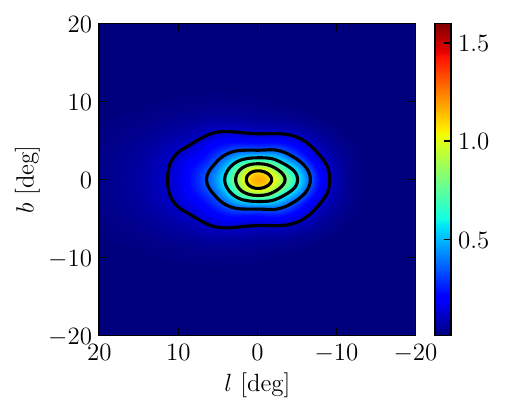}
        \includegraphics[width=0.24\columnwidth]{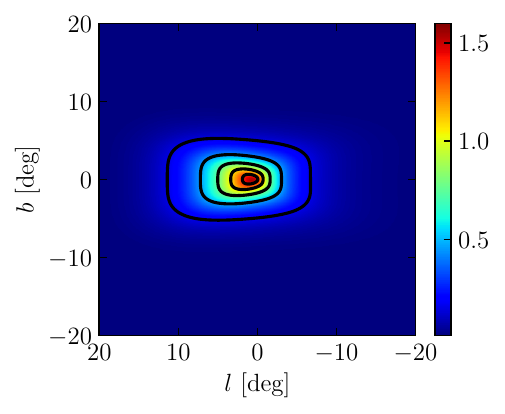}
	\caption{Bulge templates after they are convolved with the PSF at 1.02 -- 1.32 GeV. References from left to right: ~\citet{Freudenreich:1997bx},~\citet{2013MNRAS.434..595C},~\citet{Coleman:2019kax}, and~\citet{McDermott2023}.
	}
	\label{fig:bulge_templates_bin6}
\end{figure}

We repeat the analysis performed in M2023, but with additional bulge models described in Section~\ref{sec:models}.
To consistently convolve with the \textit{Fermi}-LAT PSF (which are not included in the original M2023), we run \texttt{fermitools} based on the same criteria as in~C2022 and M2023 (see Table~\ref{tab:data_cholis} for details). Namely, we use \texttt{gtselect} and \texttt{gtmktime} to select and filter the events, then use \texttt{gtbin} to bin the data. After using \texttt{gtltcube} and \texttt{gtexpcube2} to obtain the live-time and exposure, we use \texttt{gtsrcmaps} and \texttt{gtmodel} to generate the convolved templates of the three bulge models. 
In figure~\ref{fig:bulge_templates_bin6}, we display the three additional bulge templates after they are convolved with the PSF at 1.02 -- 1.32 GeV, compared with the M2023 one.

Table~\ref{tab:llh_GALPROP} reports the results using the \texttt{GALPROP}-based background model also for the different bulge templates. 
The mutual Bayes factor, $\ln \mathcal{B}_{XY} \equiv \Delta \ln \mathcal{H}$, allows us to appreciate the
performance of the different models and to assess which one performs better.
From Table~\ref{tab:llh_GALPROP}, we can see that our version of the Cao13 template 
has $\ln \mathcal{B} = 180$, when compared to the BB \texttt{gcepy} model, meaning that it provides a better model for the gamma-ray data. 
On the other hand, our other two bulge models are even better than the BB \texttt{gcepy} template, with $\ln \mathcal{B} = 507$ and 685 for F98 and Coleman20, respectively, when compared to the BB \texttt{gcepy} model. 

As for the comparison with the DM-inspired template, we see that the gNFW$^2$ is preferred over the BB \texttt{gcepy} and Cao13 model, consistent with the findings of~M2023. However, F98 has $\ln \mathcal{B} = 54$, when compared to gNFW$^2$. Finally, the Coleman20 model results to be the best GCE template in our tests for the \texttt{GALPROP}-based background model, with a $\ln \mathcal{B} = 232$, when compared to gNFW$^2$. 
We conclude, therefore, that for \texttt{GALPROP}-based models, the choice of the bulge template is crucial for interpreting whether the GCE is DM-like or bulge-like. 
We nonetheless stress that the above statement holds for the specific diffuse emission model adopted, which is not guaranteed to provide an overall good fit to gamma-ray data. 
We will discuss how ring-based models improve the goodness of fit in the next section. 

\begin{table*}
\centering
\caption{Results of the likelihood analysis for the \texttt{GALPROP}-based model using \texttt{gcepy}, including the Bayesian evidence, and mutual Bayes factor with respect to the best-fit model \texttt{GALPROP}$_{\rm 7p}$. We remind that the same mask as in M2023 is applied on 4FGL-DR2 sources and 
the Galactic plane, $|b| < 2^\circ$.
Likelihood values from \texttt{gcepy} are consistent with the ones obtained with the L-BFGS-B algorithm.}
 \label{tab:llh_GALPROP}
    \begin{tabular}{cccccc}
    \toprule
    Baseline Model & Additional source & $-2\ln{\mathcal{L}}$ & TS & $\ln{\mathcal{H}}$ & $\ln \mathcal{B} \equiv \Delta\ln{\mathcal{H}}$\\
    \midrule
    \texttt{GALPROP}$_{\rm 7p}$  & none & 3752798 & 0 & $-$1876678 & 0 \\
    \texttt{GALPROP}$_{\rm 8t}$  & BB (\texttt{gcepy}) & 3750941 & 1857 & $-$1875793 & 885 \\
    \texttt{GALPROP}$_{\rm 8t}$  & gNFW$^2$ & 3750051 & 2747 & $-$1875340 & 1338 \\
    \midrule
    \texttt{GALPROP}$_{\rm 8t}$  & Cao13 & 3750582 & 2216 & $-$1875613 & 1065 \\
    \texttt{GALPROP}$_{\rm 8t}$  & F98 & 3749924 & 2874 & $-$1875286 & 1392 \\
    \texttt{GALPROP}$_{\rm 8t}$  & Coleman20 & 3749563 & 3235 & $-$1875108 & 1570 \\
    \bottomrule
    \end{tabular}
\end{table*}

\subsection{Model systematics II: Testing ring-based models}
\label{sec:M2023-ring}
We here explore the same set of four bulge models but with the ring-based Galactic diffuse models of P2022. 
We remind the reader that M2023 also presented a run with the P2022 ring-based background model, arguing that this always provided a worse fit to the data than the \texttt{GALPROP}-based optimized background model when no excess was considered, i.e.,~\texttt{GALPROP}$_{\rm 7p}$. 

Also in this case, we run the fit with both minimizers to cross-check the validity of the \texttt{gcepy} code. 
Differently from the \texttt{GALPROP}-based models, when comparing the results of the two minimizers
for the ring-based background model, we find a major discrepancy between our results and M2023:
Contrary to the finding of M2023, when using the L-BFGS-B algorithm we find that the ring-based background model provides a better fit compared with the ${\rm \texttt{GALPROP}}_{\rm 7p}$ background model.

In Table~\ref{tab:lmfit-vs-gcepy}, we compare the likelihood results\footnote{The L-BFGS-B algorithm 
simply optimizes the likelihood 
and therefore does not allow for Bayesian evidence calculation. For this comparison, we, therefore, use $\Delta$TS values, but we will show that our conclusions also hold when considering the Bayesian evidence.} from the L-BFGS-B algorithm and the \texttt{gcepy} package. 
We focus on the no-excess case and compare the \texttt{GALPROP}-based and ring-based background models. Using the L-BFGS-B algorithm, we find that the TS for the \texttt{GALPROP}-based background model against the ring-based one is $-1852$. Using the \texttt{gcepy} package, we find instead that the \texttt{GALPROP}-based background model has a positive TS of 3508 against the ring-based one. This observation aligns qualitatively with the findings in~M2023 who report a positive TS of 4539 for the \texttt{GALPROP}-based background. 
Table~\ref{tab:lmfit-vs-gcepy} shows that the $-2\ln\mathcal{L}$ values for the \texttt{GALPROP}-based background model are almost the same between L-BFGS-B and \texttt{gcepy}. However, this is not the case for the ring-based background model.
Thus, the discrepancy is only seen in the ring-based analyses. 

\begin{figure}
    \centering
    \includegraphics[width=0.48\columnwidth]{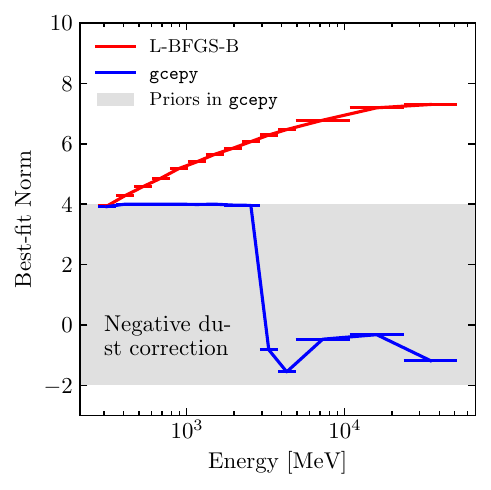}
    \includegraphics[width=0.48\columnwidth]{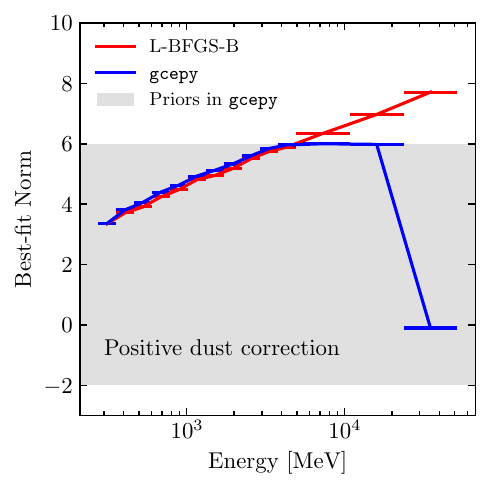}
    \caption{Best-fit normalizations for the negative and positive dust corrections from L-BFGS-B and \texttt{gcepy} (with narrow priors as in the original implementation) using the ring-based background model developed by P2022. We are plotting the dust template normalization against the 14 energy bins range from 0.275 to 51.9 GeV.}
    \label{fig:lmfit-vs-gcepy}
\end{figure}

After investigating the fitting results in each energy bin and for every template, we find that the major difference between the fits using the L-BFGS-B algorithm and \texttt{gcepy} is in the best-fit values for the negative and positive dust corrections in the ring-based model. To use nested sampling to estimate Bayesian posteriors, \texttt{gcepy} has to implement priors for the templates. The adopted priors are uniform in logarithmic space and are sufficiently wide for most templates. However, the priors for the negative and positive dust corrections turned out to be too limiting. In the public code of \texttt{gcepy}, the priors for the normalization in logarithmic space are uniform between [$-$2, 4] for the negative dust correction and [$-$2, 6] for the positive dust correction. However, when we adopted the L-BFGS-B algorithm, we provided bounds to scale the parameters from 0 to large values, far exceeding the priors in \texttt{gcepy} in linear space. Figure~\ref{fig:lmfit-vs-gcepy} shows the best-fit values for the normalization of the negative and positive dust corrections from L-BFGS-B and \texttt{gcepy} in each energy bin for the ring-based background model. It is clear that the best-fit values for the negative dust correction from L-BFGS-B always exceed the prior range of \texttt{gcepy}. In \texttt{gcepy}, the best fit found by \texttt{gcepy} simply stops at the upper boundary of the prior for most bins. For a few high-energy bins, the best fit found by \texttt{gcepy} is very small, likely caused by finding a local minimum. The same situation is also observed for the positive dust correction, although only for a few high-energy bins. 
In Table~\ref{tab:llh_lmfit}, we find that the best-fit likelihood for the ring-based model from \texttt{gcepy} is again consistent with that from the L-BFGS-B algorithm once we widen the priors for dust corrections. More specifically, we widen the prior upper bound of two dust corrections to 10 while maintaining the other priors unchanged.
We conclude that~M2023 failed to find the real best-fit models when using the ring-based background model due to inadequate priors for the dust corrections. Our results using the L-BFGS-B algorithm in Table~\ref{tab:lmfit-vs-gcepy}, and the \texttt{gcepy} results with wider priors in  Table~\ref{tab:llh_lmfit}, provide a more accurate interpretation of the ring-based background model.

\begin{table*}
\centering
\caption{Comparison between the L-BFGS-B algorithm and \texttt{gcepy} with its original priors for the \texttt{GALPROP}$_{\rm 7p}$-based background model and the ring-based background model without GCE. 
} 
\label{tab:lmfit-vs-gcepy}
    \begin{tabular}{cccc}\toprule
    Background model & Fitting algorithm & $-2\ln\mathcal{L}$ & $2\left(\ln{\cal L}_{{\texttt{GALPROP}}_{\rm 7p}}-\ln\cal{L}_{\rm ring-based} \right)$\\\midrule
    \texttt{GALPROP}$_{\rm 7p}$ & \multirow{2}{*}{L-BFGS-B} & 3752792 & \multirow{2}{*}{$-$1852} \\
    ring-based &  & 3750940 & \\\hline
    \texttt{GALPROP}$_{\rm 7p}$ & \multirow{2}{*}{\texttt{gcepy}} & 3752798 & \multirow{2}{*}{$+$3508} \\
    ring-based &  & 3756306 & \\\bottomrule
    \end{tabular}
\end{table*}

\begin{figure}
    \centering
    \includegraphics{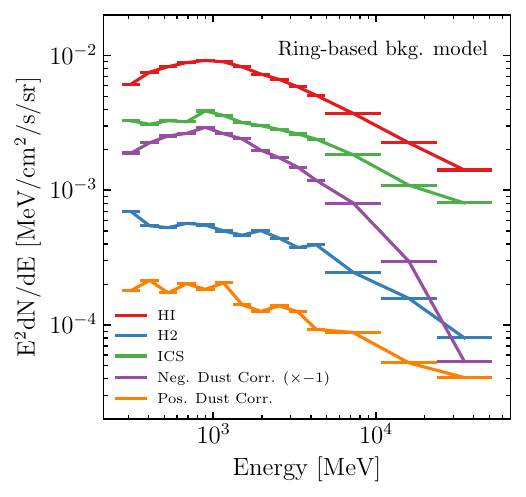}
    \caption{Best-fit spectra for the HI, H2, ICS, negative and positive dust corrections for the ring-based background model (P2022). The results are obtained with no GCE template and are from the L-BFGS-B algorithm.}
    \label{fig:spectra_gas_residuals}
\end{figure}

In the 'no-excess' case, as detailed in Figure~\ref{fig:spectra_gas_residuals}, the best-fit spectra include the four HI and H2 rings and six ICS rings, along with both negative and positive dust corrections. Notably, the negative dust correction at GeV energies is about 30\% of the total HI and H2 fluxes, while the positive correction is relatively small. 
As the negative dust correction template represents an underestimation of the spin temperature, we would expect 
a constant ratio of negative dust correction to the HI spectrum across all energies.
Although, this appears to mainly be the case, a deviation occurs at the highest energy bin. However, this discrepancy is not crucial for determining the GCE's morphology, as the GCE's values are minimal at such high energies.

\begin{figure}
    \centering
    \includegraphics[width=0.48\columnwidth]{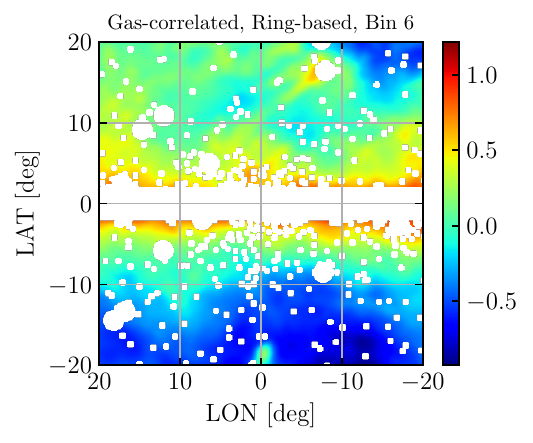}
    \includegraphics[width=0.48\columnwidth]{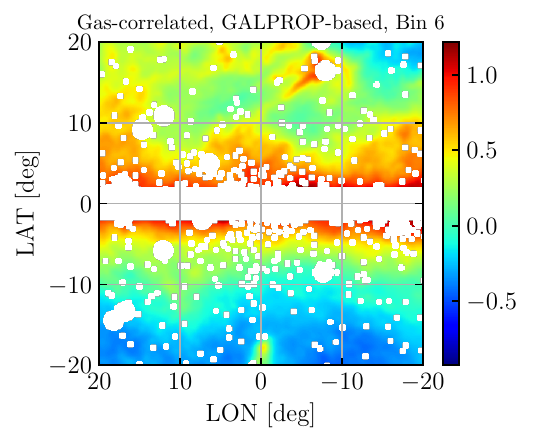}
    \caption{Best-fit count maps (in $\log_{10}$ scale) for the gas-correlated component for the ring-based and \texttt{GALPROP}$_{\rm 7p}$ based background model. We show the 6th energy bin (1.02--1.32 GeV).}
    \label{fig:cmaps}
\end{figure}

Figure~\ref{fig:cmaps} displays the best-fit count map (in $\log_{10}$ scale) for the gas-correlated component (HI, H2, and the dust corrections) in the ring-based background model at the 1.02--1.32 GeV energy bin. For comparison, we also show the gas-correlated component (pion decay + bremsstrahlung) in the best-fit \texttt{GALPROP}-based background model for the same energy bin. Due to the negative correction, the photon counts associated with the gas-correlated component in the ring-based model are generally lower than those in the \texttt{GALPROP}-based model. However, no pixels in the unmasked 
region have negative values when HI, H2, and dust corrections are combined. We have verified that this holds true for every energy bin.

In Table~\ref{tab:llh_lmfit}, 
 we add the results from using the ring-based background model using wider priors in gcepy. In this case, again, likelihood values with the alternative minimizer are comparable.
 Comparing row one of Table~\ref{tab:llh_GALPROP} with row one of
Table~\ref{tab:llh_lmfit} shows that the ring-based model without GCE is a better description of the
gamma-ray sky than the \texttt{GALPROP}$_{\rm 7p}$, with a $\ln \mathcal{B} = 216$.
When adding a GCE component, regardless of the choice of the template, the Bayesian evidence for the GCE is overall reduced for the ring-based background model with respect to  \texttt{GALPROP}-based background models, cf.~Table~\ref{tab:llh_GALPROP} and Table~\ref{tab:llh_lmfit}. 
Yet the evidence of the GCE, no matter the template adopted, is strong, i.e.~$\ln \mathcal{B} \gtrsim 100$. 
The ranking of GCE models is similar to Table~\ref{tab:llh_GALPROP}, 
except for the fact of F98 performing worse than Cao13 ($\ln \mathcal{B} = 26$). The gNFW$^2$ template is still preferred over Cao13 ($\ln \mathcal{B} = 44$), while Coleman20 template provides a better fit than the gNFW$^2$ template ($\ln \mathcal{B} = 88$). We, therefore, corroborate the Coleman20 preference found in the \texttt{GALPROP}-based runs, even when the ring-based background model is used. 

We notice, however, that the overall goodness of fit of models with the GCE and using the \texttt{GALPROP}$_{\rm 8t}$ background template are preferred with respect to our optimization of the
ring-based background runs with $\ln \mathcal{B}$ varying between about 500 and 1000 for the different GCE templates. 
Nonetheless, we will show below, Section~\ref{sec:igrb}, that these very same models lead to unphysical 
spectra of the IGRB, questioning their physical interpretations.

\begin{figure}
    \centering
    \includegraphics[width=0.48\columnwidth]{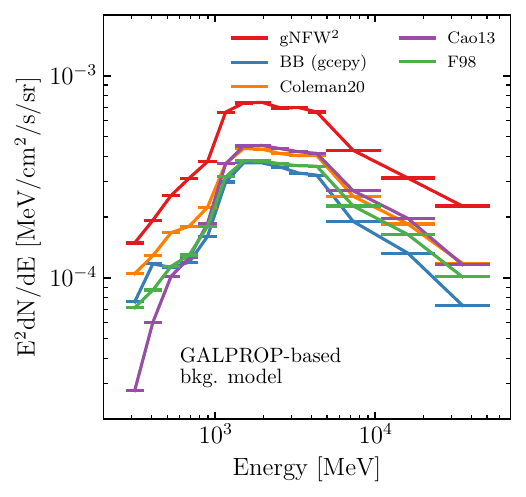}
    \includegraphics[width=0.48\columnwidth]{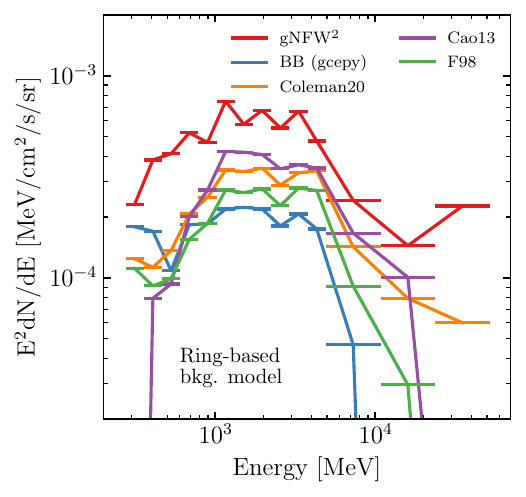}
    \caption{GCE spectra for the gNFW$^2$ template as well as the four bulge templates from different references, using the \texttt{GALPROP}$_{8t}$-based background model (left) and the  ring-based background model (right).}
    \label{fig:gce-spectra}
\end{figure}

Figure~\ref{fig:gce-spectra} shows the GCE spectra for the five templates tested using both the \texttt{GALPROP}-based and the ring-based background models. Overall, the GCE fluxes are higher when using the gNFW$^2$ template compared with bulge templates, by a factor of a few. All the GCE spectra are relatively soft, and their 
$E^2 dN/dE$
values peak at around 1--2 GeV.

\begin{table*}
\centering
  \caption{
  Similar to Table \ref{tab:llh_GALPROP} except here we consider the ring-based background model developed by P2022. We have expanded the priors for the dust corrections in \texttt{gcepy} to ensure convergence. Note that in each row, the amplitude of the ring-based background model templates is optimized in addition to the GCE additional source (if there is one) to maximize the likelihood $\cal L$, as explained in Section~\ref{sec:models}.  }
 \label{tab:llh_lmfit}
    \begin{tabular}{cccccc}
    \toprule
    Baseline Model & Additional source & $-2\ln \mathcal{L} $ & TS & $\ln \mathcal{H}$ & $ \ln{\mathcal{B}}\equiv\Delta\ln{\mathcal{H}}$\\
    \midrule
    ring-based  & none & 3750994 & 0 & $-$1876462 & 0 \\
    ring-based  & BB (\texttt{gcepy}) & 3750592 & 402 & $-$1876297 & 165 \\
    ring-based  & F98 & 3750570 & 424 & $-$1876302 & 160 \\
    ring-based  & Cao13 & 3750560 & 434 & $-$1876276 & 186 \\
    ring-based  & gNFW$^2$ & 3750433 & 561 & $-$1876232 & 230 \\
    ring-based  & Coleman20 & 3750333 & 661 & $-$1876144 & 318 \\
    \bottomrule
    \end{tabular}
\end{table*}

\subsection{Model systematics: degeneracy between IGRB and GCE}
\label{sec:igrb}
\begin{figure}
    \centering
    \includegraphics[width=0.32\columnwidth]{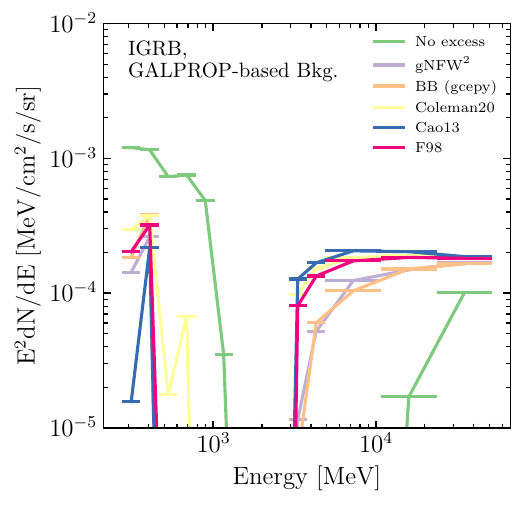}
    \includegraphics[width=0.32\columnwidth]{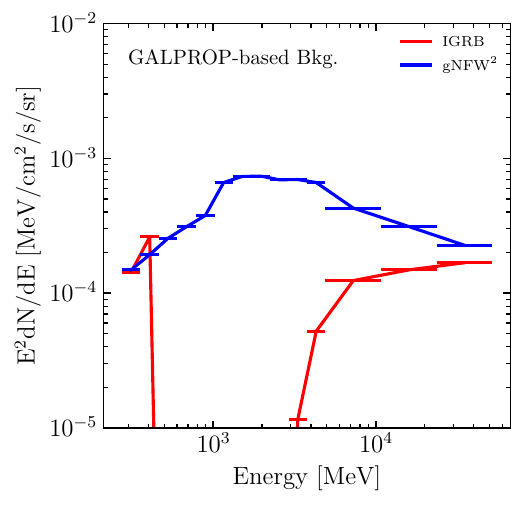}
    \includegraphics[width=0.32\columnwidth]{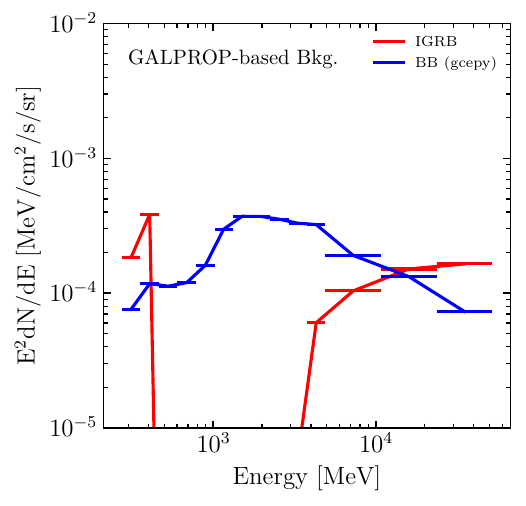}
    \includegraphics[width=0.32\columnwidth]{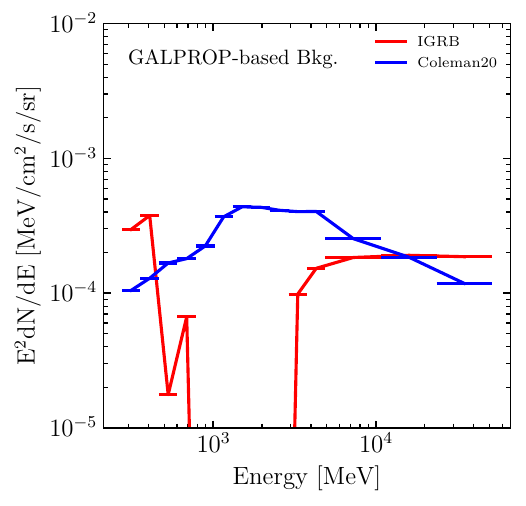}
    \includegraphics[width=0.32\columnwidth]{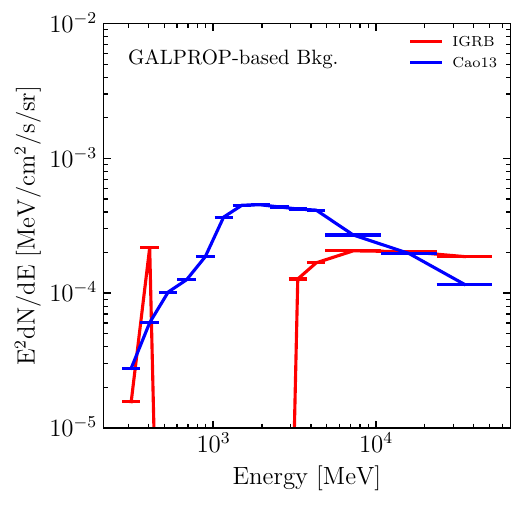}
    \includegraphics[width=0.32\columnwidth]{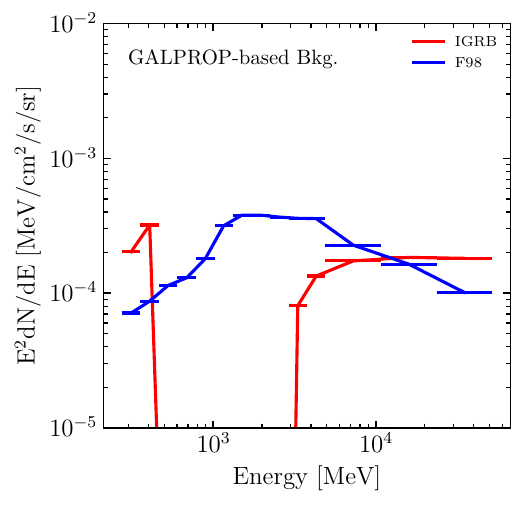}
    \caption{\emph{Top left.} Spectra of IGRB in the \texttt{GALPROP}-based background model, using different GCE templates (including no excess). \emph{Remaining.} Comparing the IGRB and GCE fluxes when different GCE templates are used in the \texttt{GALPROP}-based background model.}
    \label{fig:igrb-gce-GALPROP}
\end{figure}

\begin{figure}
    \centering
    \includegraphics[width=0.32\columnwidth]{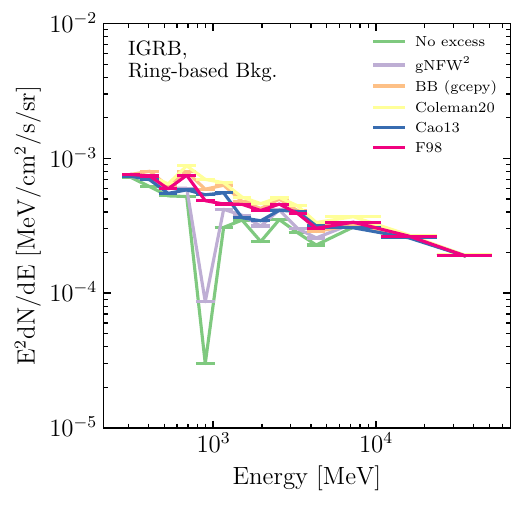}
    \includegraphics[width=0.32\columnwidth]{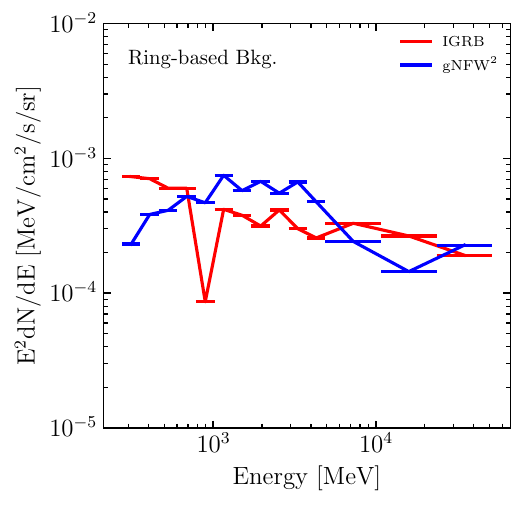}
    \includegraphics[width=0.32\columnwidth]{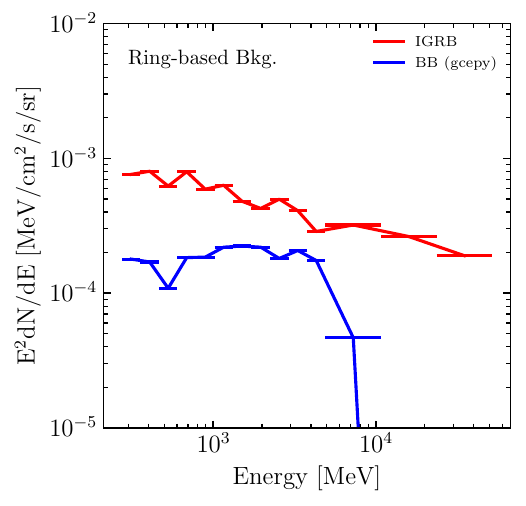}
    \includegraphics[width=0.32\columnwidth]{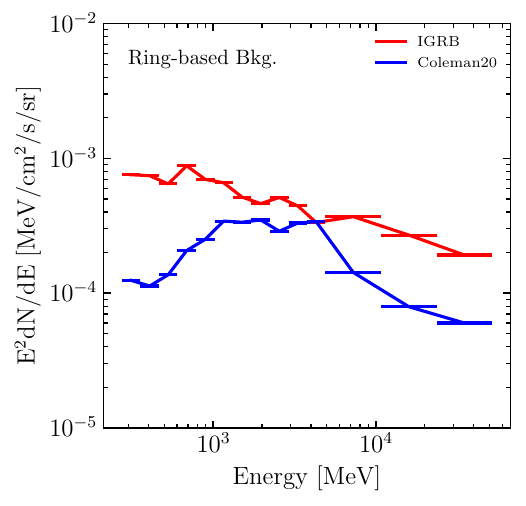}
    \includegraphics[width=0.32\columnwidth]{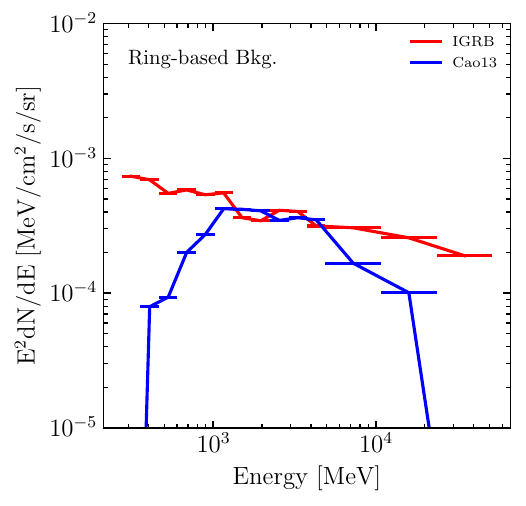}
    \includegraphics[width=0.32\columnwidth]{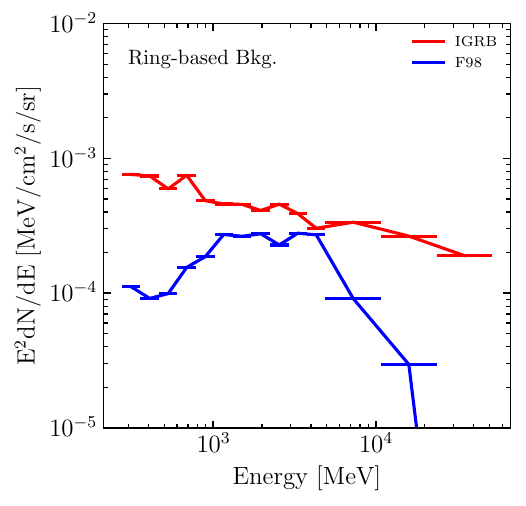}
    \caption{Same as figure~\ref{fig:igrb-gce-GALPROP}, but for the ring-based background model.}
    \label{fig:igrb-gce-rings}
\end{figure}

In Figure~\ref{fig:igrb-gce-GALPROP} and~\ref{fig:igrb-gce-rings} we examine the spectra for the IGRB for the different background models and GCE templates. 
As can be seen from the top left panel of Figure~\ref{fig:igrb-gce-GALPROP}, when using the \texttt{GALPROP}-based background model, instead, the IGRB is hardly ``detected'' around GeV energies, no matter which GCE template is used (including no excess case).
This result seems to be  unphysical.
Comparing the IGRB and GCE flux for different GCE templates (remaining panels of Figure~\ref{fig:igrb-gce-GALPROP}), the IGRB is largely missing around where the GCE is peaked. The issue does not seem to exist for the ring-based background model,
which is shown in Figure~\ref{fig:igrb-gce-rings}.
The fact that the IGRB in Figure~\ref{fig:igrb-gce-GALPROP} has no flux around 1--5 GeV
is a concerning property of the \texttt{GALPROP}-based background models, which is not seen in the ring-based models.
One possible way to ameliorate this issue may be to put strong priors on the IGRB flux so it doesn't drop to zero.

\subsection{Analysis systematics: The role of mask size and goodness of fit with Monte Carlo simulations}
\label{sec:Masked Analysis Using Data-Driven Diffuse Galactic Emission  Templates}
\begin{figure}
	\centering
	\includegraphics{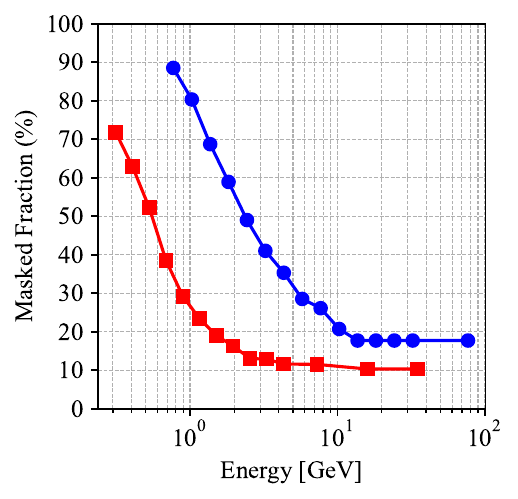}
	\caption{A comparison of the fraction of the sky masked for the analysis in of the sky masked in Section~\ref{sec:Masked Analysis Using Data-Driven Diffuse Galactic Emission  Templates} (blue)
 and the fraction of sky masked for the rest of Section~3 and also M2023 (red).
  Note that in both cases, the plot is for the combined point source and Galactic plane mask.
	}
	\label{fig:mask_size}
\end{figure}

For the sake of studying the systematic uncertainties induced by the choice of the masked regions, we use the same data and templates as P2022. The main difference with the ring-based case in Section~\ref{sec:M2023-ring} is that the pixel resolution used is now $0.2^\circ$ rather than $0.1^\circ$. We also use a larger point-source mask, the details of which are shown in Table~\ref{tab:point_source_mask}. The new point-source mask was designed to be wide enough to mask out 90\% of the flux of each point source in each energy bin.  Our masks can be compared to Table~1 of the supplementary material of M2023.  As can be seen, our point-source masks have larger radii than even their “large” point-source masks. We found that if we make our masks smaller, then we could see the residual point source signal leaking out from the mask.  Note that we do not use the first two energy bins listed in Table~\ref{tab:point_source_mask} in our subsequent analysis for this subsection, as those bins have too high a fraction of the ROI masked out to provide meaningful constraints.
In Figure~\ref{fig:mask_size}, we compare the fraction of the ROI that is masked in this section to the fraction of the ROI that is masked in earlier sections and in M2023. We also remind the reader that in M2023 and in previous sections, the Galactic plane was masked.

\begin{table}
\centering
\begin{tabular}{|c|c|c|c|}
\toprule
$E_{\rm min}$--- $E_{\rm max}$ [GeV]  & $\theta$[$^\circ$] & Masked fraction \\
\midrule
0.667 --- 0.889 & 1.92 & 88.5\% \\
0.889 --- 1.19 & 1.58 & 80.3\% \\
1.19 --- 1.58 & 1.28 & 68.7\% \\
1.58 --- 2.11 & 1.04 & 58.9\% \\
2.11 --- 2.81 & 0.8 & 49.0\% \\
2.81 --- 3.75 & 0.72 & 41.0\% \\
3.75 --- 5.0 & 0.56 & 35.3\% \\
5.0 --- 6.67 & 0.48 & 28.5\% \\
6.67 --- 8.89 & 0.36 & 26.1\% \\
8.89 --- 11.9 & 0.32 & 20.7\% \\
11.9 --- 15.8 & 0.2 & 17.7\% \\
15.8 --- 21.1 & 0.2 & 17.7\% \\
21.1 --- 28.1 & 0.2 & 17.7\% \\
28.1 --- 37.5 & 0.2 & 17.7\% \\
37.5 --- 158 & 0.2 & 17.7\% \\
\bottomrule
\end{tabular}
\caption{The energy bins and the radii,  $\theta$, of the point source masks used for Section~\ref{sec:Masked Analysis Using Data-Driven Diffuse Galactic Emission  Templates}.
		 The last column shows the fraction of pixels masked relative to the total number of pixels in the inner $40^{\circ} \times 40^{\circ}$ GC region, including both the 4FGL-DR2 catalogue masked point sources and the Galactic plane $\left|b\right| < 2^\circ$ mask.}
\label{tab:point_source_mask}
\end{table}

In Table~\ref{tab:loglike-values}, we evaluate the statistical evidence for the additional GCE component,
when our new point-source mask, as described above, is applied (but the Galactic plane is unmasked). 
Since we do not mask the Galactic plane, we also add a model component for the 
NB \citep{Nishiyama2015}.
This table can be compared to table 2 of P2022. As can be seen, the qualitative results are very similar to P2022, with mild difference in the likelihoods probably due to the masking and also to having discarded the first two energy bins so that we have 13 bins rather than the 15 that were used by P2022. As can be seen from the table, we find that, with a larger point-source mask, there is strong evidence for an additional component on top of the ring-based background model. Moreover, there is evidence for the Coleman20 template at 
8.1$\sigma$ on top of the ring-based+NB model, while the addition of 
gNFW$^2$ is not significant (2.8$\sigma$).
Finally, the significance for DM is strongly reduced to a negligible level when added to the ring-based+NB+Coleman model. 

We then run the case with both new point-source mask and Galactic plane mask $\left|b\right|< 2^\circ$. 
In this case, the results are shown in Table~\ref{tab:loglike-values-noplane}.
As can be seen, with the Galactic plane and and new point-source mask, we find neither the gNFW$^2$ DM template nor the NB template to be significant. Conversely, the Coleman20 BB template is still significant.

\begin{table}
\centering
\caption{{ Statistical significance of the GCE templates for the ring-based  background model of P2022 when the new point-source mask is applied (plane unmasked).} Additional sources considered in the analysis are NB, Coleman20 BB, and gNFW$^2$ DM-like template. In addition to the TS, the significance of the additional component is also given in terms of the equivalent number of $\sigma$.
}\label{tab:loglike-values}
\begin{tabular}{cccc}
\toprule
Baseline model  & Additional source &  $\mathrm{TS}$  &  Significance \\ 
\midrule
ring-based  & Coleman20            &  77.5        & $7.3\;\sigma$\\ 
 ring-based  & gNFW$^2$             & 80.7        & $7.5\;\sigma$\\ 
 ring-based  & NB                & 299.7       & $16.2\;\sigma$\\ \hline
 ring-based+NB  & gNFW$^2$               &  21.0     & $2.8\;\sigma$\\ 
 ring-based+NB  & Coleman20                &  90.9 &$8.1\;\sigma$\\\hline 
 ring-based+NB+Coleman20  & gNFW$^2$              &    3.5    & $0.3\;\sigma$\\ 
\bottomrule
\end{tabular}
\end{table}

\begin{table}
\centering
\caption{The same as Table~\ref{tab:loglike-values} except that the Galactic plane ($\left|b\right|< 2^\circ$) is also masked.
}\label{tab:loglike-values-noplane}
\begin{tabular}{cccc}
\toprule
Baseline model  & Additional source &  $ \mathrm{TS}$  &  Significance \\ 
                \hline
  ring-based  & gNFW$^2$              & 12        & $1.7\;\sigma$\\ 
 ring-based  & NB                & 19       & $2.6\;\sigma$\\ 
  ring-based  & Coleman20             &  56        & $5.9\;\sigma$\\ 
 \hline
 ring-based+Coleman20   & NB               &  3     & $0.2\;\sigma$\\ 
 ring-based+Coleman20   & gNFW$^2$                &  5 &$0.5\;\sigma$\\
\bottomrule
\end{tabular}
\end{table}

\begin{figure}
    \centering
    \includegraphics[width=0.45\columnwidth]{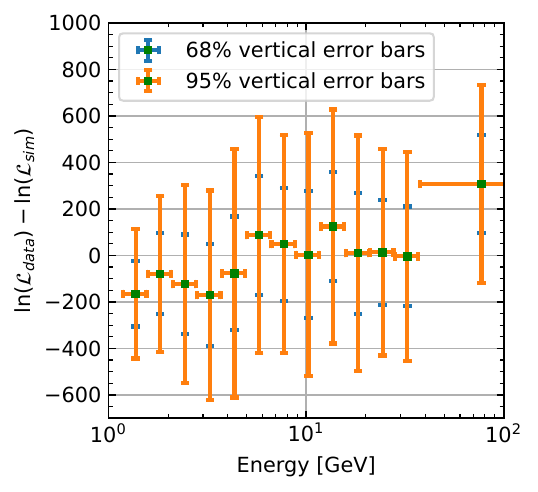}
    \includegraphics[width=0.45\columnwidth]{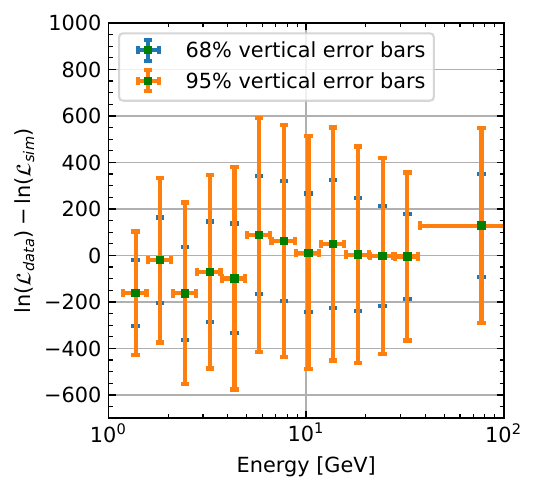}
    \caption{{\em Left}: 
    	Monte Carlo simulations for the ring-based+NB+Coleman20 model when the point sources are masked as specified in Table~\ref{tab:point_source_mask}.
    	Poissonian simulations were generated for the best-fit parameters of this model. Each simulation was fit using the ring-based+NB+Coleman20 model, and the maximum likelihood for the simulation ($\mathcal{L}_{\rm sim}$) was compared to the maximum likelihood for the  ring-based+NB+Coleman20 model fit to the \textit{Fermi}-LAT data ($\ln \mathcal{L}_{\rm data}$) which is given in Table~\ref{tab:loglike-values}. The vertical error bars were estimated from the mean and standard deviation of the simulation samples. The horizontal error bars indicate the energy bin widths.
     {\em Right}: Same as the left plot except that the Monte Carlo simulations are for the ring-based+Coleman20 model when the point sources are masked as specified in Table~\ref{tab:point_source_mask} and the $\left|b\right|< 2^\circ$ Galactic plane was also masked. }
    \label{fig:MC_sims_M2_B2}
\end{figure}

P2022 tested the goodness of fit using Monte Carlo simulations. As can be seen from their Figure~9, the Monte Carlo simulations were not consistent with the fit for the $E<5$~GeV. This was somewhat ameliorated to $E<4$~GeV by reducing the ROI from 40$^\circ$ by 40$^\circ$ to 30$^\circ$ by 30$^\circ$. In Fig.~\ref{fig:MC_sims_M2_B2}, 
we show the full ROI Monte Carlo simulations for the new point-source mask, and for the case with both the point sources and Galactic plane masked out respectively. As can be seen, in all energy bins, the Monte Carlo simulations are consistent with the data. This indicates that, with standard template fitting, it is more robust to mask the point sources rather than try to model them when the diffuse Galactic emission is being fit. 
An alternative would be to include a model of the point sources and simultaneously fit the position of the point sources with the parameters of the Galactic diffuse emission model. However, this would be very computationally intensive and goes beyond the scope of tests necessary in the current analysis.

\section{Adaptive-template fitting}
\label{sec:skyfact_analysis}

In this section, we show that the potential of \texttt{skyFACT} to reduce residuals and optimize model components in a data-driven way allows for robust inference on the GCE morphology, as it was already shown in the case of the analysis of sub-threshold point sources~\citep{Calore:2021jvg}. Due to a large number of nuisance parameters, it is infeasible to optimize a given gamma-ray emission model on small ROIs, in particular, ROIs with a masked Galactic plane that encompasses the bulk of the detected gamma rays from the GC. Whenever we optimize a model with \texttt{skyFACT}, we, therefore, perform the optimization with respect to the full ROI of $40^{\circ}\times40^{\circ}$ centred on the GC.

In the \texttt{skyFACT}-based part of this study, we work with the \textit{Fermi}-LAT data selected according to Table~\ref{tab:data_cholis} except for a change in bin size from $0.1^{\circ}$ to $0.25^{\circ}$ to render the analysis computationally tractable. 
We consider three distinct model compositions in parallel, namely the \texttt{GALPROP}$_{\rm 8t}$, ring-based, and the original \texttt{skyFACT}  \citep{Bartels:2017vsx} backgrounds.
In all three gamma-ray emission model setups, we employ the Coleman20, NB and gNFW$^2$ templates. Note that in this section, we only consider masking the Galactic plane. The point sources are added to the background model with full spectral freedom per point source. In the following section, we outline how we apply the \sky~re-modulation in the context of a masked ROI.

\subsection{Deriving gamma-ray optimized background models with \sky}
\label{sec:skyfact_A}

\sky~enables us to go beyond the model iterations investigated in Section~\ref{sec:M2023} by re-modulating the spatial morphology of the respective model's components (where possible). As stated earlier in Section~\ref{sec:skyfact-technique}, this is achieved via adaptive template fitting, invoking a large number of spectral (per energy bin) and spatial modulation parameters (per spatial pixel) whose ranges are controlled by user-input hyper-parameters. The degree of variation in these modulation parameters is restricted via a penalizing likelihood function adding to a standard Poisson likelihood term to prevent overfitting.

Such an approach is only feasible with enough information in the considered dataset. Thus, applying \sky~in the presence of an extensive Galactic plane mask is prohibitive. Therefore, we devise the following scheme to incorporate re-modulated gamma-ray emission models in our analysis: For each of the considered background model setups stated above, we perform a fit to the \textit{Fermi}-LAT data -- data selection described in Table~\ref{tab:data_cholis} -- by enabling spatial re-modulation. \sky~hyperparameter settings are given in Section~\ref{sec:skyfact-technique}. We obtain what we call ``\textit{optimized}'' versions of the original background model setups. \sky's optimization will re-modulate the background templates to minimize the residual photons, i.e.~parts of the GCE emission will be absorbed by the selection of background components. We deem such an approach conservative because we deliberately diminish the total luminosity of the GCE. However, since the GCE's spatial morphology and spectrum are not fully degenerate with the employed background components, it is very unlikely that the entire excess is re-absorbed in the optimized background templates (and indeed, we will confirm this with our results). In what follows, we investigate the performance of these optimized background model iterations on datasets with a Galactic plane mask utilizing, in a second step, standard template fits.

To convey an idea of how the \sky~optimized models compare to the original versions, we explicitly go through the derivation of the optimized version of P2022. In Table~\ref{tab:model_setup_Pohletal}, we list the astrophysical gamma-ray emission model components of the original setup of P2022, which we employ in this part of the analysis. The selected spatial templates and associated spectra guarantee the optimization of each component based on physical priors. The GCE components, in particular, are initialised following the average spectrum of Galactic MSPs detected by \textit{Fermi}-LAT \citep{McCann:2014dea}, which suppresses the gamma-ray emission above a few tens of GeV.\footnote{Further exploration of the GCE properties above 10 GeV with \texttt{skyFACT} is presented in \cite{Manconi:2024DMlimits}.}

\begin{figure}
    \centering

\includegraphics[width=0.49\columnwidth]{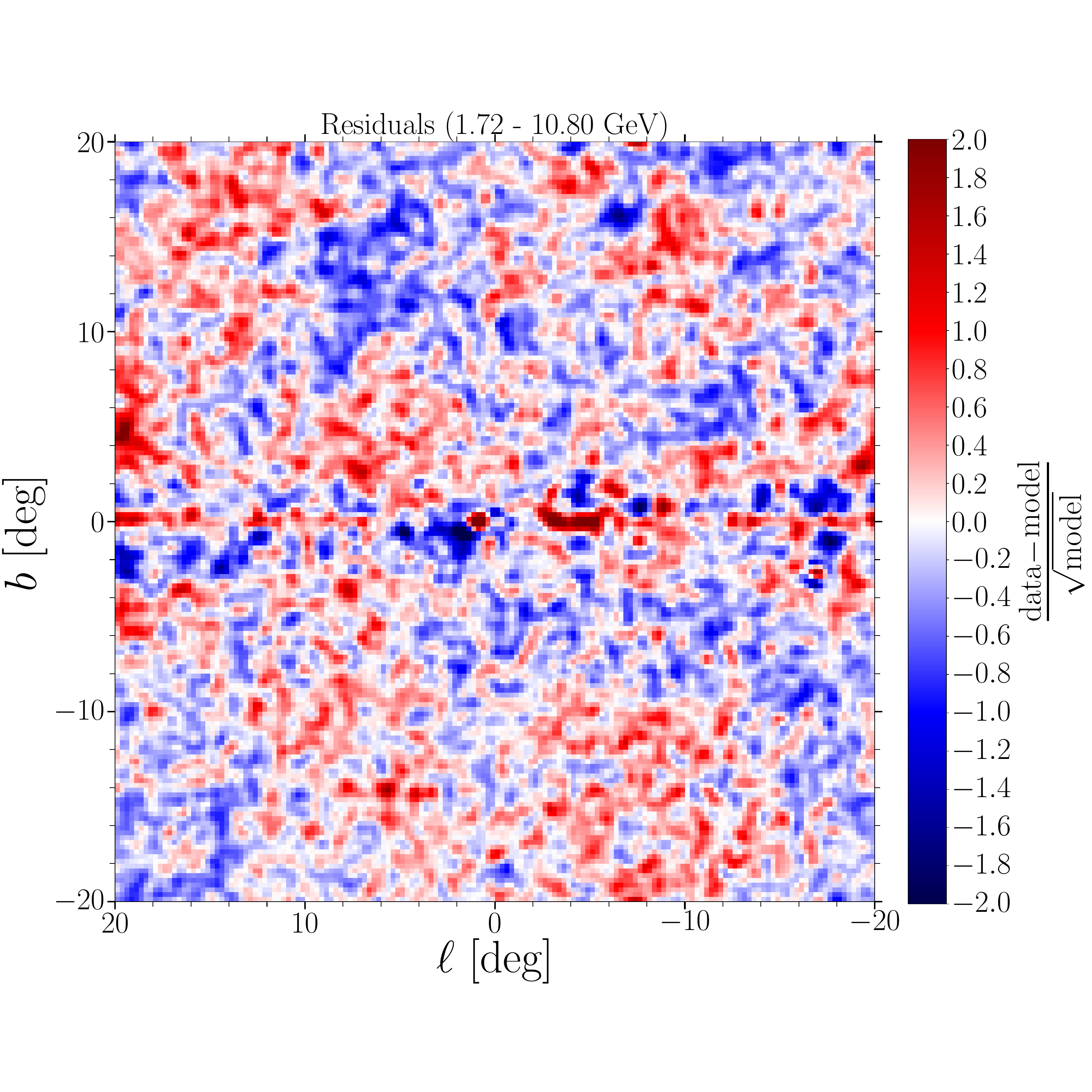}
    \includegraphics[width=0.49\columnwidth]{
    	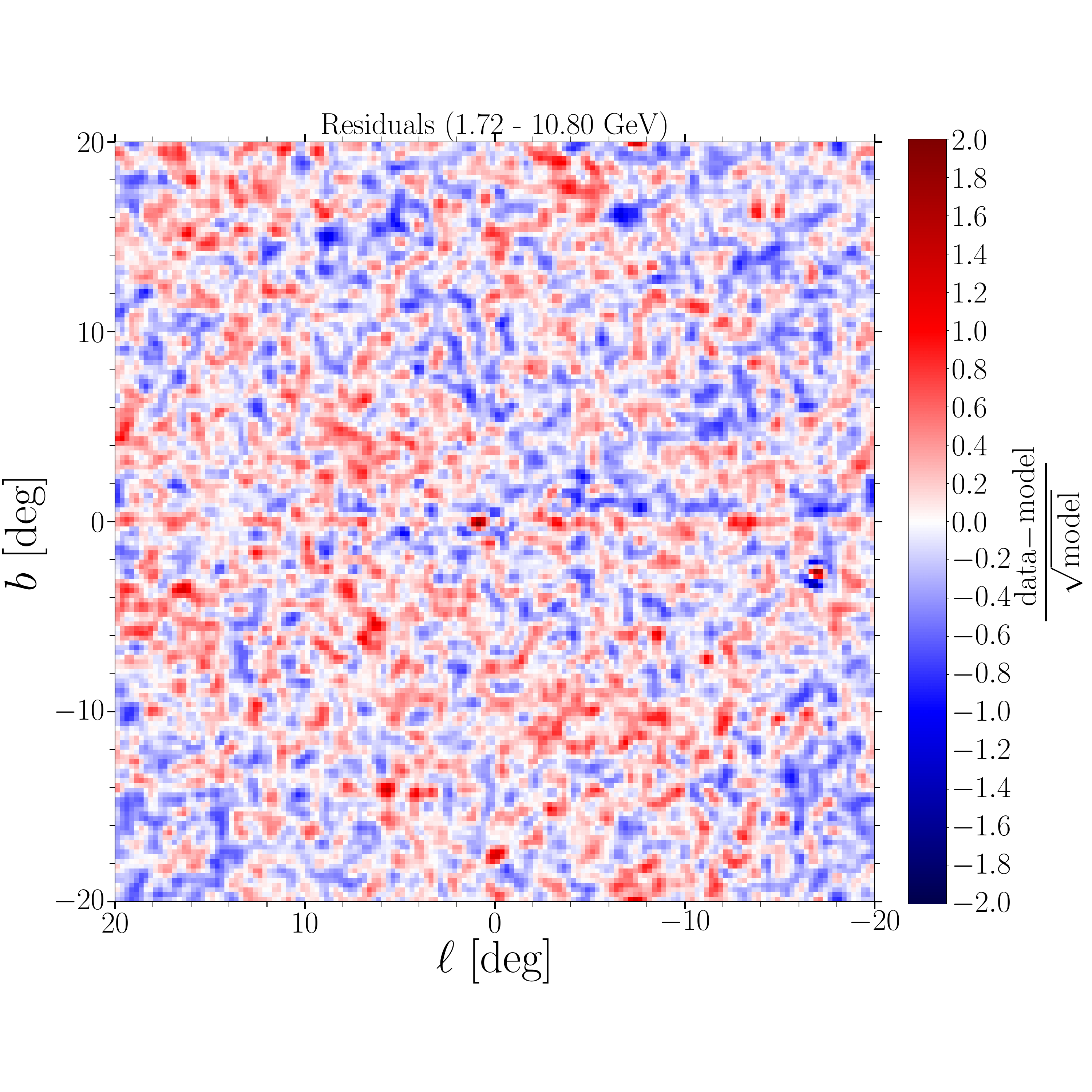}
    \caption{Shows $(\mathrm{data} - \mathrm{model}) / \sqrt{\mathrm{model}}$ of residuals in the integrated energy range from 1.72 GeV to 10.8 GeV. (\emph{Left}:) Employing \texttt{skyFACT} to perform a template fit with the ring-based astrophysical model components listed in Table~\ref{tab:model_setup_Pohletal} with a GCE represented by the Coleman20 + NB + gNFW$^2$. (\emph{Right}:) Running the same model setup with the full re-modulation power of \texttt{skyFACT} to optimize the employed components.}
    \label{fig:residuals_objectiveA}
\end{figure}

To reduce the required computation time of \texttt{skyFACT} and to improve its convergence, we combine the HI and H2 templates per ring into a single component, and we create a single inverse Compton template from the six initial rings. The results of the optimization run -- in this case using the example of modeling the GCE with Coleman20, NB, and gNFW$^2$ to exemplify what is maximally possible regarding the reduction of fit residuals -- are shown in the right panel of Figure~\ref{fig:residuals_objectiveA} in terms of significance ($ (\mathrm{data} - \mathrm{model}) / \sqrt{\mathrm{model}}$) of the residuals in the energy range from 1.72 GeV to 10.8 GeV. We compare these residuals to the residuals we obtain with a simple template fit (left panel of the same figure) using the full model as described in Table~\ref{tab:model_setup_Pohletal}, and which should correspond to results in Section~\ref{sec:M2023-ring}. Here, a template fit refers to a maximum likelihood fit where the spatial morphology of all templates is fixed, i.e.~turning off all spatial and spectral re-modulation parameters, while all spectra are completely unconstrained and free to vary. As intended, \texttt{skyFACT} is able to noticeably reduce the significant residuals of the template fit along the Galactic plane to the left and right of the GC. Moreover, the remaining residuals of the optimized model appear rather featureless and well-distributed around zero. We obtain very similar results when the GCE is modelled with only the gNFW$^2$ DM template. This optimization procedure is repeated for the remaining background models, M2023's \texttt{GALPROP}$_\mathrm{8t}$ and the model setup of \texttt{run5} used in the original \sky~works.

\begin{table*}
\centering
  \caption{Summary of the spatial and spectral model components used as input for \texttt{skyFACT} to derive an optimized version of the gamma-ray emission model of P2022}
 \label{tab:model_setup_Pohletal}
 \begin{tabular}{lcc}
  \toprule
component & spatial morphology & spectrum\\ 
\hline
\multirow{2}{*}{$\pi^0$ + bremsstrahlung} & HI maps P2022 and H2 maps \cite{Macias:2016nev}
in four rings  & \multirow{2}{*}{\cite{Fermi-LAT:2012edv}}\tabularnewline
 & (in Galactocentric radii: 0--3.5 kpc, 3.5--8 kpc, 8--10 kpc,
10--50 kpc)  & \tabularnewline
\hline
\multirow{2}{*}{inverse Compton} & numerical computation in six Galactocentric rings via \texttt{\texttt{GALPROP} v56}  & spectrum of foreground \tabularnewline
 & based on the 3D interstellar radiation field model of \citep{Porter:2017vaa}  & model A \citep{Fermi-LAT:2014ryh} \tabularnewline
\hline
dust correction  & positive and negative corrections maps of \cite{Fermi-LAT:2019yla} & power law $\propto E^{-2}$ \\
\hline
detected sources &  4FGL-DR2 sources in our ROI \citep{Ballet:2020hze} & spectra listed in 4FGL-DR2 \\
\hline
isotropic gamma-ray background & isotropic & Fermi Science Tools \\
\hline
\multirow{2}{*}{Fermi Bubbles} & \multirow{2}{*}{\cite{Macias:2019omb}} & \cite{Fermi-LAT:2017opo} \tabularnewline 
 & & for low latitudes\\
\hline
Sun and Moon  & data-driven, derived with the Fermi Science Tools & Fermi Science Tools\\
\hline
Loop I  & \citep{2007ApJ...664..349W} & power law $\propto E^{-2}$\\
\hline
DM  & C2022 &  $\propto (E/1 \,\mathrm{GeV})^{-1.46}\exp{(-E/3.6\,\mathrm{GeV})}$\\
\hline
Boxy Bulge  & \cite{Coleman:2019kax} & $\propto (E/1 \,\mathrm{GeV})^{-1.46}\exp{(-E/3.6\,\mathrm{GeV})}$\\
\hline
Nuclear stellar cluster  & \cite{Nishiyama2015} & $\propto (E/1 \,\mathrm{GeV})^{-1.46}\exp{(-E/3.6\,\mathrm{GeV})}$\\
\bottomrule
 \end{tabular}
\end{table*}

\subsection{Dark matter evidence in masked analyses with the original \texttt{skyFACT} setup.} 
\label{sec:skyfact_B}

To investigate the preferred spatial morphology of the GCE, we turn towards an alternative gamma-ray emission model neither probed in P2022 nor M2023. The \texttt{run5} model iteration compiled for the original \sky~works \citep{Storm:2017arh, Bartels:2017vsx} provides an ideal candidate to shed light on the impact of a Galactic plane mask and its impact on template-based fits. To this end, we fully adopt the setup of \texttt{run5} in terms of model composition (see the cited publications for all details) and \sky's hyperparameter settings. This gamma-ray emission model iteration contains representatives of most of the components listed in Table~\ref{tab:model_setup_Pohletal} except for the dust correction, Loop I, Sun and Moon contributions.

In contrast to the other two background model iterations, this one can only be used in its ``optimized'' version as the original templates of the model are not meant to fit the data well. Following our rationale outlined in the previous section, we first perform an optimization run with respect to the selected \textit{Fermi}-LAT dataset without masking any part of the sky while adding no explicit GCE components to the model definition. Afterwards, we extract the optimized model components with the aim of conducting several template fits based on the optimized background templates with varying Galactic plane mask sizes. In these template fits, we restrict the full energy range of the selected \textit{Fermi}-LAT dataset to energy bins covering 500 MeV to 12 GeV, i.e.~10 energy bins in total, in order to save computation time while still capturing the bulk of the GCE's emission.

To derive a statistically sound assessment of the data's preference for any particular GCE morphology, we need to perform template fits of nested models. We perform this type of model comparison in terms of the significance of the additional component as explained in Section~\ref{sec:statistical-framework}.
We define our base model as all astrophysical background components plus a GCE represented by Coleman20 and NB templates. An extended model adds the gNFW$^2$ template so that we can compare the likelihood values for fits with both model instances. 
Adding a gNFW$^2$ component to the base model essentially adds 11 degrees of freedom or parameters, which must be strictly positive, namely the normalizations of the DM template per energy bin and a global normalisation parameter.

The results of this 
approach are reported in Table~\ref{tab:objectiveB_significances}.
The first row of this table can be compared to the last row of Table 2 of P2022. There a slightly lower significance was found but some minor differences are to be expected given variations on how the point sources are treated, the number of energy bins used, and the modulation done of the background templates. In both cases, the significance of a gNFW$^2$ template is negligible once the Coleman20 BB and NB have been added.
The results of Table~\ref{tab:objectiveB_significances} indicate that the evidence for the necessity of adding a gNFW$^2$ template to the gamma-ray emission model is at most $1.6\sigma$ in the case of no Galactic plane mask. Interpreted differently, there is only marginal evidence that the preferred morphology of the GCE follows a gNFW$^2$ profile. 
This is in agreement with results of Section~\ref{sec:Masked Analysis Using Data-Driven Diffuse Galactic Emission  Templates}, also when spatial modulation of the background model components is allowed.

\begin{table}
    \centering
    \caption{\label{tab:objectiveB_significances} Summary of the significance
    	 of an additional gNFW$^2$ template in template fits with the \sky~\texttt{run5} setup for various Galactic plane mask sizes. Here ``base'' refers to the \sky~optimized \texttt{run5} background templates plus the Coleman20 BB and a NB component. The \sky~optimization of background templates was performed on the \texttt{run5} setup without GCE components as outlined in Sec.~\ref{sec:skyfact_A}.}
    \begin{tabular}{lcccc}
    \toprule
    Galactic plane mask & $-2\ln \mathcal{L}_{\mathrm{base+gNFW^2}}$ & $-2\ln\mathcal{L}_{\mathrm{base}}$ & TS & significance \\
    \hline
no mask & 275081 & 275071 & 10 & 1.6$\sigma$ \\ 
$|b|<1^{\circ}$ & 260989 & 260985 & 4 & 0.5$\sigma$ \\ 
$|b|<2^{\circ}$ & 247030 & 247029 & 1 & 0.1$\sigma$ \\
$|b|<5^{\circ}$ & 205937 & 205932 & 5 & 0.7$\sigma$ \\
    \bottomrule
    \end{tabular}
\end{table}

\subsection{Bayesian model comparison of original and optimized gamma-ray emission models}
\label{sec:skyfact_C}

Given the sometimes opposing findings on the preferred spatial morphology of the GCE reported in the broad body of literature, it is necessary to ask the question of how much the employed astrophysical background model impacts the eventual conclusion. Here, we investigate this question from a Bayesian point of view by quantifying the degree of belief in certain gamma-ray emission models, that is, what model fits the \textit{Fermi}-LAT data best. The expectation is to verify that with increasing Bayesian evidence for a gamma-ray emission model, the preference for a particular spatial morphology of the GCE is converging to either DM represented by a gNFW$^2$ template or the combination of the Coleman20 BB and NB templates. 

\paragraph*{Model comparison without GCE components.} In Table~\ref{tab:objectiveC_backgrounds}, we consider six background models, three original background model template sets and three optimized template sets obtained by applying the rationale outlined in Section~\ref{sec:skyfact_A}. We first compare the performance of these models in a template fit without adding additional GCE components, i.e.~we assess how well the background templates alone fit the GC gamma-ray emission. Note that we also include M2023's \texttt{GALPROP}$_{\rm 7p}$ background model (only original templates) since it was found in M2023 that it yields the best fit when not accounting for GCE components. In all runs, we applied a Galactic plane mask of $|b|<2^{\circ}$, which turned out to be a crucial ingredient in the comparisons of DM-like and bulge templates.

To derive the stated Bayesian evidence $\mathcal{H}$, we proceed as follows: We extract the best-fitting template normalizations for each model iteration. We sum all model components multiplied by the retrieved best-fitting normalizations. Using \texttt{MultiNest}~\citep{Feroz:2008xx} and specifying 1000 live points and an evidence tolerance of 0.2, we re-fit the masked ROI with these models while assigning a single normalization parameter to it and employing a Poisson log-likelihood function.

By comparing these models,\footnote{We stress that the numbers presented in Table~\ref{tab:objectiveC_backgrounds} cannot and should not be directly compared to the previous results in Section~\ref{sec:M2023}. The main reason is the differing resolution of the templates adopted for the two analyses. In Section~\ref{sec:M2023}, we use a finer resolution of $0.1^{\circ}$, which directly translates to larger likelihood values. At the same time, we have to project the gamma-ray flux models of each component to the chosen geometry of the dataset. The initial resolution of the flux model is most of the times coarser than the chosen bin size for the data. Projection effects can distort and washout information so that we do not compare the exact same model each time we change the resolution.} we notice that the optimized versions of each respective background model iteration always yield a better fit to the data, also in the case of ring-based background models. In contrast to M2023, we do not find that the original \texttt{GALPROP}$_{\rm 7p}$ performs better than the original \texttt{GALPROP}$_{\rm 8t}$ without additional GCE components as shown by a Bayes factor of $\ln \mathcal{B} = 277$ in favour of the latter.

Among all considered model iterations, we find that the  \texttt{run5} model of the original \sky~works yields the best description of the data by far ($\ln \mathcal{B} > 1000$ to the next best iteration, the optimized M2023's \texttt{GALPROP}$_{\rm 8t}$). Yet, already at this stage, we caution the reader not to over-interpret the quoted evidence values $-\ln\!{\mathcal{H}}$. \sky~is not a perfect tool and cannot re-modulate the templates to 100\% accuracy. For example, when we compare the optimized templates associated with the $\pi^0$ and bremsstrahlung emission (following the gas distribution in the Milky Way) among different background model iterations, we find that they do not converge to the exact same morphology. For instance in the most extreme case, the relative deviation of this optimized gas-related component is on average around 30\% with respect to \sky's \texttt{run5} and M2023's \texttt{GALPROP}$_{\rm 8t}$. On one side, this is caused by the quite diverse gamma-ray emission model composition in P2022, M2023 and \sky's \texttt{run5}, which yields different priors for the adaptive template fitting routine. On the other side, this technique relies on user-defined hyperparameters that alter the final results. While we improve the fit results via \sky, we do not claim to have derived the unique optimal diffuse model describing the physics of the GC.

\paragraph*{Model comparison with added stellar GCE components.} For each of the all gamma-ray emission models in Table~\ref{tab:objectiveC_backgrounds} but \texttt{GALPROP}$_{\rm 7p}$ original, we then add an additional GCE component, modeled as 
Coleman20 and NB, perform a standard template fit and extract Bayesian evidence values as described in the previous paragraph. 
As can be seen from Table~\ref{tab:objectiveC_BB_significances}, in general, we find very strong evidence for the combination of the Coleman20 BB and NB on top of the background-only model iterations, as clearly implied by the large significance values around 20$\sigma$ in all tested cases. This means that, regardless of the \sky~optimization, {\it Fermi}-LAT data strongly want an additional component, i.e.~the GCE. 
As can be seen from Table~\ref{tab:objectiveC_BB_significances}, among the ``original'' non-\sky~modulated gamma-ray emission models, we find that the one proposed by M2023, \texttt{GALPROP}$_{\rm 8t}$,  exhibits strong evidence ($\ln{B} \approx 300$) for being a better fit to the data with respect to the original P2022 setup. This claim has been made in M2023, which we are able to reproduce here (and in Section~\ref{sec:M2023}). However, this does not mean that their model is, in general, the best description of reality since we are only looking at latitudes $|b| > 2^\circ$. The \sky-modulated versions of both gamma-ray emission model instances are strongly preferred by the data compared to their original counterparts. Globally, the \sky~\texttt{run5} model is the one performing best among all tested cases, even when spatial modulation are allowed on the original M2023 and P2022 models. There is strong evidence for it being preferred over the second-best model, M2023 optimized, by $\ln \mathcal{B}\approx 1400$.
  
\paragraph*{Model comparison and significance of a DM component.} Finally, we added a gNFW$^2$ component in the subsequent fit, on top of the bulge Coleman20 and NB model. This way, we can repeat the approach outlined in Section~\ref{sec:fitting-and-statistics} to quantify the statistical significance of an additional gNFW$^2$ template from the frequentist perspective.
 As can be seen from Table~\ref{tab:objectiveC_DM_significances}, we are also able to reproduce the M2023 result of the strong evidence (more than $11\sigma$) for the necessity of an additional DM gNFW$^2$ template on top of Coleman20 and NB in the context of the original M2023 model setup. However, as can also be seen from Table~\ref{tab:objectiveC_DM_significances}, the significance of the gNFW$^2$ component is only marginal after \sky~modulating the spatial morphology of the background model. 
In contrast, consistent with earlier works and the results in Table~\ref{tab:loglike-values-noplane}, we see in Table~\ref{tab:objectiveC_DM_significances} that the setup of P2022 never required an additional gNFW$^2$ template after accounting for the GCE as the Coleman20 BB template and NB template. 
As can also be seen from Table~\ref{tab:objectiveC_DM_significances}, the \sky~model is an outlier here because the fit, including a gNFW$^2$ template, is even worse than the one without. This situation can occur in \sky~since even in a template fit, we modulate the spatial morphology of the detected extended 4FGL-DR2 sources. Thus, the penalizing likelihood function adds a non-vanishing part to the overall value of the likelihood function. Yet, the extensions of these sources are marginal compared to the rest of the ROI, so we do not expect biased results. Consequently, there is also no evidence of the need for a gNFW$^2$ template in this gamma-ray emission model iteration.

In conclusion, whenever we employ an optimized astrophysical background model, there is no strong evidence for spherical symmetry of the GCE or at least a preference for such a morphology even when masking the Galactic plane. Previous contrary findings seem to be driven by a certain amount of background mismodelling. We stress again that quoted values of the Bayesian evidence are subject to the caveats raised in Sec.~\ref{sec:statistical-framework}. Eventually, the Bayesian framework allowed us to single out the \sky~\texttt{run5} setup to be the most suitable to describe the Galactic center physics with adaptive template fitting. Its assumed priors for the spatial and spectral profile of the used components yield the best-fit model among the tested cases although it does by no means imply that it is the optimal model achievable.

\begin{table}
	\centering
	\caption{\label{tab:objectiveC_backgrounds} Summary of the likelihoods ($\cal L$) and evidence ($\mathcal{H}$) for different background models . A Galactic plane mask of $|b|<2^{\circ}$ is applied, but the point sources are included as part of the background model. The optimised version of the background model includes a \sky~modulated version of the non-point source components of the ``original'' model. }
\begin{tabular}{lrr}
	\toprule
	Background model &  $-2\ln(\mathcal{L})$ & $-\ln\!{\mathcal{(H)}}$\\
	\midrule
	M2023's \texttt{GALPROP}$_{\rm 7p}$ "original"&          347477 & 280175\\
	M2023's \texttt{GALPROP}$_{\rm 8t}$ "original"&          347465 & 279898\\
	M2023's \texttt{GALPROP}$_{\rm 8t}$ "optimized" &          342008 & 274036\\
	P2022's ring-based "original" &          346859 & 279723\\
	P2022's ring-based "optimized" &          342982 & 276075 \\
	\sky~\texttt{run5} "optimized" &          340266 & 272900\\
	\bottomrule
\end{tabular}
\end{table}

\begin{table}
	\centering
	\caption{\label{tab:objectiveC_BB_significances} Summary of the TSs, significances  for the Coleman20+NB templates  for different background models. The evidence ($\cal H$) is also given for the combined model of background  + Coleman20 + NB.
		See Table~\ref{tab:objectiveC_backgrounds} for an explanation of the background models and their corresponding likelihoods.
  }
\begin{tabular}{lrrr}
	\toprule
	Background model &  TS(Coleman20+NB) &  significance &  $-\ln({\cal H})$ \\
	\midrule
	M2023's \texttt{GALPROP}$_{\rm 8t}$ "original" &     2137 &   > 20$\sigma$ &     278929 \\
	M2023's \texttt{GALPROP}$_{\rm 8t}$ "optimized" &     4822 &   > 20$\sigma$ &     272018 \\
	P2022's ring-based "original" &      445 &   19.1$\sigma$ &     279231 \\
	P2022's ring-based "optimized" &     5435 &   > 20$\sigma$ &     272573 \\
	\sky~\texttt{run5} "optimized" &     3957 &   > 20$\sigma$ &     270627 \\
	\bottomrule
\end{tabular}
\end{table}

\begin{table}
	\centering
	\caption{\label{tab:objectiveC_DM_significances} Summary of the TSs and significances for the gNFW$^2$ template  for different background models once the Coleman20+NB templates have been added.
		See Table~\ref{tab:objectiveC_backgrounds} for an explanation of the background models and their corresponding likelihoods.}
\begin{tabular}{lrlr}
	\toprule
	Base model &  TS(gNFW$^2$) &  significance & $-\ln\!{(\mathcal{H})}$ \\
	\midrule
	M2023's \texttt{GALPROP}$_{\rm 8t}$ "original" + Coleman20+NB &    162 &  11.4$\sigma$ & 278843\\
	M2023's \texttt{GALPROP}$_{\rm 8t}$ "optimized" + Coleman20+NB &     22 &   2.7$\sigma$ &  272008\\
	P2022's ring-based "original" + Coleman20+NB &     32 &   3.9$\sigma$ & 279202\\
	P2022's ring-based "optimized" + Coleman20+NB &      14 &   1.7$\sigma$ & 272504\\
	\sky~\texttt{run5} "optimized" + Coleman20+NB&    -3&     -- & 270673 \\
	\bottomrule
\end{tabular}
\end{table}

\section{Conclusions}
\label{sec:conclusions}
We have performed an extensive analysis of models of gamma-ray emission towards the GC as an explanation of the GCE, 
subject to different choices of diffuse background models, point source and Galactic plane masking, and extended source models. 
In particular, we tested \texttt{GALPROP}-based background models vs.~more flexible non-parametric
ring-based models.

First, we have thoroughly tested contradicting results in the literature for masked analyses, especially those pertaining to the preference for stellar bulge vs.~DM, e.g., in M2023, where preference for a gNFW$^2$ (DM-like) emission of the GCE was found. 
In Section~\ref{sec:M2023}, we showed that we can reproduce the analysis in M2023, and we scrutinized their main results in light of model and analysis systematic uncertainties. 
In Section~\ref{sec:M2023-bulge}, we highlighted the relevance of the bulge templates: when using the same \texttt{GALPROP}-based background models as in M2023, 
with the same data selection and masks, we demonstrated using Bayesian evidence that the Coleman20 and F98 bulge models provide a better description
of the inner Galaxy gamma-ray sky than a gNFW$^2$ 
model.
We then tested an alternative model for the Galactic diffuse emission, and, in particular, 
the so-called ring-based background model. In Section~\ref{sec:M2023-ring}, we demonstrated that, 
in the absence of an additional GCE source, the ring-based model better performs with respect
to \texttt{GALPROP}-based models. 
When adding a GCE source, the Coleman20 bulge model is the preferred model of the GCE, significantly better than the gNFW$^2$ template.
We notice that the contrary conclusions of M2023 when using the ring-based model were due to their not correctly finding the minimum of the parameters for the ring-based templates due to an overly restrictive prior. We also found that M2023 used a non-standard version of the Galactic bulge template. This is confirmed by the fact that even with their \texttt{GALPROP}-based templates, the better-motivated Coleman20 bulge template still provides a superior fit to the \textit{Fermi}-LAT data.
In Section~\ref{sec:Masked Analysis Using Data-Driven Diffuse Galactic Emission  Templates}, we examined the case of ring-based templates with more aggressive masking. We found that, when just the point sources are masked out, the Coleman20 BB and NB significantly improve the fit, and once they are added, the gNFW$^2$ template does not significantly improve the fit anymore. When the Galactic plane is masked out, only the Coleman20 BB template is required. The NB is indeed too small in its spatial extent to have any significant effect on the model fit to the data in that case. We also showed using Monte Carlo simulations that the fits were consistent with simulations. 

We then looked at the case where templates could be spatially modulated using \sky. Allowing for more freedom on the spatial parts of the model components, we were able to further minimize the residuals and improve the goodness of fit.
We compared the different background models from the previous sections and added an additional \sky~model (\texttt{run5}). We demonstrated that switching on the spatial modulation of background models always provide a better fit to data, because of reduction of the residuals.
Among all considered background model iterations, we find that the \texttt{run5} model of the original \sky~works yielded the best description of the data by far ($\ln \mathcal{B} > 1000$ to the next best iteration, the optimized M2023's \texttt{GALPROP}$_{\rm 8t}$). Nonetheless, limitations still exist in the current \sky~implementation. While we improve the fit results via \sky, we do not claim to have derived the unique optimal diffuse model describing the physics of the GC. 
No matter what the background model is, we always found strong evidence for the Coleman20+NB model.
Moreover, for all background optimized models there is no additional
evidence for a DM-like signal. We found similar results, i.e.~DM evidence on top of the bulge model always below the 4$\sigma$ threshold, for almost all background models. We encountered one exception, namely the original M2023 \texttt{GALPROP}-based template. However, this was found to have a much lower Bayesian evidence in comparison to the \sky~\texttt{run5} model.
Finally, we found that, for the \texttt{run5} model of the original \sky~implementation, the evidence for an additional DM-like contribution is not significant on top of the Coleman20 bulge and NB model regardless of the cut on Galactic latitude.

We stress that throughout this work we have adopted Bayesian statistics when performing model comparison, and our conclusions have to be interpreted in such a statistical framework.

In summary, 
the preference for a bulge-like morphology of the GCE in the various analyses we have done puts on even more solid grounds the possibility that part of the excess originates from unresolved point sources, such as millisecond pulsars. 
Future multi-wavelength analyses of the GC will help determine the nature of the sources emitting across the multi-messenger spectrum from radio~\citep{Calore:2015bsx}, X-rays~\citep{Berteaud:2020zef}, up to very-high-energy gamma rays~\citep{Song:2019nrx, Macias:2021boz}.

\bigskip

{ Note: While our article was near completion, a new article came out 
\citep{ZhongCholis24}
which found that when the \texttt{GALPROP}-based background model was used, the Coleman20 bulge had a similar likelihood to the gNFW$^2$. No Bayesian model comparison is performed therein.}

\section*{Acknowledgements}
We thank the authors of M2023 and Roland Crocker for helpful discussions, and the authors of M2023 for making their analysis tools public. 
We also thank Silvia Manconi for constructive feedback on the manuscript.
The work of SH is supported by the U.S.~Department of Energy Office of Science under award number DE-SC0020262, NSF Grant No.~AST1908960 and No.~PHY-2209420, and JSPS KAKENHI Grant Number JP22K03630 and JP23H04899. The work of MK and KA is supported by NSF Grant No.~PHY-2210283. This work was supported by World Premier International Research Center Initiative (WPI Initiative), MEXT, Japan. FC and CE acknowledge support by the ``Agence Nationale de la Recherche'', grant n.~ANR-19-CE31-0005-01 (PI: F.~Calore). The work of CE has been supported by the EOSC Future project which is co-funded by the European Union Horizon Programme call INFRAEOSC-03-2020, Grant Agreement 101017536. CE further acknowledges support from the COFUND action of Horizon Europe’s Marie Sk\l{}odowska-Curie Actions research programme, Grant Agreement 101081355 (SMASH). The work of DS is supported by JSPS KAKENHI Grant Number 20H05852.

\section*{Data Availability}

The templates and some of the software used in this analysis may be obtained by contacting the authors.

\bibliographystyle{mnras}
\bibliography{morphology_GCE} 

\begin{thebibliography}{}
\makeatletter
\relax
\def\mn@urlcharsother{\let\do\@makeother \do\$\do\&\do\#\do\^\do\_\do\%\do\~}
\def\mn@doi{\begingroup\mn@urlcharsother \@ifnextchar [ {\mn@doi@}
  {\mn@doi@[]}}
\def\mn@doi@[#1]#2{\def\@tempa{#1}\ifx\@tempa\@empty \href
  {http://dx.doi.org/#2} {doi:#2}\else \href {http://dx.doi.org/#2} {#1}\fi
  \endgroup}
\def\mn@eprint#1#2{\mn@eprint@#1:#2::\@nil}
\def\mn@eprint@arXiv#1{\href {http://arxiv.org/abs/#1} {{\tt arXiv:#1}}}
\def\mn@eprint@dblp#1{\href {http://dblp.uni-trier.de/rec/bibtex/#1.xml}
  {dblp:#1}}
\def\mn@eprint@#1:#2:#3:#4\@nil{\def\@tempa {#1}\def\@tempb {#2}\def\@tempc
  {#3}\ifx \@tempc \@empty \let \@tempc \@tempb \let \@tempb \@tempa \fi \ifx
  \@tempb \@empty \def\@tempb {arXiv}\fi \@ifundefined
  {mn@eprint@\@tempb}{\@tempb:\@tempc}{\expandafter \expandafter \csname
  mn@eprint@\@tempb\endcsname \expandafter{\@tempc}}}

\bibitem[\protect\citeauthoryear{Abazajian}{Abazajian}{2011}]{Abazajian:2010zy}
Abazajian K.~N.,  2011, \mn@doi [J. Cosmol. Astropart. Phys.]
  {10.1088/1475-7516/2011/03/010}, 03, 010

\bibitem[\protect\citeauthoryear{Abazajian \& Kaplinghat}{Abazajian \&
  Kaplinghat}{2012}]{Abazajian:2012pn}
Abazajian K.~N.,  Kaplinghat M.,  2012, \mn@doi [Phys. Rev. D]
  {10.1103/PhysRevD.86.083511}, 86, 083511

\bibitem[\protect\citeauthoryear{Abazajian, Canac, Horiuchi  \&
  Kaplinghat}{Abazajian et~al.}{2014}]{Abazajian:2014fta}
Abazajian K.~N.,  Canac N.,  Horiuchi S.,   Kaplinghat M.,  2014, \mn@doi
  [Phys. Rev. D] {10.1103/PhysRevD.90.023526}, 90, 023526

\bibitem[\protect\citeauthoryear{Abazajian, Horiuchi, Kaplinghat, Keeley  \&
  Macias}{Abazajian et~al.}{2020}]{Abazajian:2020tww}
Abazajian K.~N.,  Horiuchi S.,  Kaplinghat M.,  Keeley R.~E.,   Macias O.,
  2020, \mn@doi [Phys. Rev. D] {10.1103/PhysRevD.102.043012}, 102, 043012

\bibitem[\protect\citeauthoryear{Abdollahi et~al.}{Abdollahi
  et~al.}{2020}]{Fermi-LAT:2019yla}
Abdollahi S.,  et~al., 2020, \mn@doi [Astrophys. J. Suppl.]
  {10.3847/1538-4365/ab6bcb}, 247, 33

\bibitem[\protect\citeauthoryear{Acero et~al.}{Acero
  et~al.}{2016}]{Fermi-LAT:2016zaq}
Acero F.,  et~al., 2016, \mn@doi [] {10.3847/0067-0049/223/2/26}, 223, 26

\bibitem[\protect\citeauthoryear{Ackermann et~al.}{Ackermann
  et~al.}{2012}]{Fermi-LAT:2012edv}
Ackermann M.,  et~al., 2012, \mn@doi [Astrophys. J.]
  {10.1088/0004-637X/750/1/3}, 750, 3

\bibitem[\protect\citeauthoryear{Ackermann et~al.}{Ackermann
  et~al.}{2015}]{Fermi-LAT:2014ryh}
Ackermann M.,  et~al., 2015, \mn@doi [Astrophys. J.]
  {10.1088/0004-637X/799/1/86}, 799, 86

\bibitem[\protect\citeauthoryear{Ackermann et~al.}{Ackermann
  et~al.}{2017}]{Fermi-LAT:2017opo}
Ackermann M.,  et~al., 2017, \mn@doi [Astrophys. J.]
  {10.3847/1538-4357/aa6cab}, 840, 43

\bibitem[\protect\citeauthoryear{Ajello et~al.}{Ajello
  et~al.}{2016}]{TheFermi-LAT:2015kwa}
Ajello M.,  et~al., 2016, \mn@doi [Astrophys. J.] {10.3847/0004-637X/819/1/44},
  819, 44

\bibitem[\protect\citeauthoryear{Atwood et~al.}{Atwood
  et~al.}{2009}]{Fermi-LAT:2009ihh}
Atwood W.~B.,  et~al., 2009, \mn@doi [Astrophys. J.]
  {10.1088/0004-637X/697/2/1071}, 697, 1071

\bibitem[\protect\citeauthoryear{Ballet, Burnett, Digel  \& Lott}{Ballet
  et~al.}{2020}]{Ballet:2020hze}
Ballet J.,  Burnett T.~H.,  Digel S.~W.,   Lott B.,  2020, \mn@doi [arXiv]
  {10.48550/arXiv.2005.11208}, \href
  {https://ui.adsabs.harvard.edu/abs/2020arXiv200511208B} {p. arXiv:2005.11208}

\bibitem[\protect\citeauthoryear{Bartels, Krishnamurthy  \& Weniger}{Bartels
  et~al.}{2016}]{Bartles2016}
Bartels R.,  Krishnamurthy S.,   Weniger C.,  2016, \mn@doi [Phys. Rev. Lett.]
  {10.1103/PhysRevLett.116.051102}, 116, 051102

\bibitem[\protect\citeauthoryear{Bartels, Storm, Weniger  \& Calore}{Bartels
  et~al.}{2018}]{Bartels:2017vsx}
Bartels R.,  Storm E.,  Weniger C.,   Calore F.,  2018, \mn@doi [Nat. Astron.]
  {10.1038/s41550-018-0531-z}, 2, 819

\bibitem[\protect\citeauthoryear{Berteaud, Calore, Clavel, Serpico, Dubus  \&
  Petrucci}{Berteaud et~al.}{2021}]{Berteaud:2020zef}
Berteaud J.,  Calore F.,  Clavel M.,  Serpico P.~D.,  Dubus G.,   Petrucci
  P.-O.,  2021, \mn@doi [Phys. Rev. D] {10.1103/PhysRevD.104.043007}, 104,
  043007

\bibitem[\protect\citeauthoryear{Binney, Gerhard, Stark, Bally  \&
  Uchida}{Binney et~al.}{1991}]{BinneyUnderstandingkinematicsGalactic1991}
Binney J.,  Gerhard O.~E.,  Stark A.~A.,  Bally J.,   Uchida K.~I.,  1991, Mon.
  Not. R. Astron. Soc., \href
  {https://ui.adsabs.harvard.edu/abs/1991MNRAS.252..210B} {252, 210}

\bibitem[\protect\citeauthoryear{{Bland-Hawthorn} \& Gerhard}{{Bland-Hawthorn}
  \& Gerhard}{2016}]{Bland-Hawthorn2016}
{Bland-Hawthorn} J.,  Gerhard O.,  2016, \mn@doi [Ann. Rev. of A \& A]
  {10.1146/annurev-astro-081915-023441}, \href
  {http://adsabs.harvard.edu/abs/2016ARA\%26A..54..529B} {54, 529}

\bibitem[\protect\citeauthoryear{Buschmann, Rodd, Safdi, Chang, Mishra-Sharma,
  Lisanti  \& Macias}{Buschmann et~al.}{2020}]{Buschmann:2020adf}
Buschmann M.,  Rodd N.~L.,  Safdi B.~R.,  Chang L.~J.,  Mishra-Sharma S.,
  Lisanti M.,   Macias O.,  2020, \mn@doi [Phys. Rev. D]
  {10.1103/PhysRevD.102.023023}, 102, 023023

\bibitem[\protect\citeauthoryear{Calore, Cholis  \& Weniger}{Calore
  et~al.}{2015}]{Calore:2014xka}
Calore F.,  Cholis I.,   Weniger C.,  2015, \mn@doi [J. Cosmol. Astropart.
  Phys.] {10.1088/1475-7516/2015/03/038}, 03, 038

\bibitem[\protect\citeauthoryear{Calore, Di~Mauro, Donato, Hessels  \&
  Weniger}{Calore et~al.}{2016}]{Calore:2015bsx}
Calore F.,  Di~Mauro M.,  Donato F.,  Hessels J. W.~T.,   Weniger C.,  2016,
  \mn@doi [Astrophys. J.] {10.3847/0004-637X/827/2/143}, 827, 143

\bibitem[\protect\citeauthoryear{Calore, Donato  \& Manconi}{Calore
  et~al.}{2021}]{Calore:2021jvg}
Calore F.,  Donato F.,   Manconi S.,  2021, \mn@doi [Phys. Rev. Lett.]
  {10.1103/PhysRevLett.127.161102}, 127, 161102

\bibitem[\protect\citeauthoryear{Cao, Mao, Nataf, Rattenbury  \& Gould}{Cao
  et~al.}{2013}]{2013MNRAS.434..595C}
Cao L.,  Mao S.,  Nataf D.,  Rattenbury N.~J.,   Gould A.,  2013, \mn@doi
  [MNRAS] {10.1093/mnras/stt1045}, \href
  {https://ui.adsabs.harvard.edu/abs/2013MNRAS.434..595C} {434, 595}

\bibitem[\protect\citeauthoryear{Caron, Eckner, Hendriks, J{\'o}hannesson,
  {Ruiz de Austri}  \& Zaharijas}{Caron et~al.}{2023}]{Caron:2022akb}
Caron S.,  Eckner C.,  Hendriks L.,  J{\'o}hannesson G.,  {Ruiz de Austri} R.,
   Zaharijas G.,  2023, \mn@doi [J. Cosmol. Astropart. Phys.]
  {10.1088/1475-7516/2023/06/013}, 06, 013

\bibitem[\protect\citeauthoryear{Chang, Mishra-Sharma, Lisanti, Buschmann, Rodd
   \& Safdi}{Chang et~al.}{2020}]{Chang:2019ars}
Chang L.~J.,  Mishra-Sharma S.,  Lisanti M.,  Buschmann M.,  Rodd N.~L.,
  Safdi B.~R.,  2020, \mn@doi [Phys. Rev. D] {10.1103/PhysRevD.101.023014},
  101, 023014

\bibitem[\protect\citeauthoryear{Cholis, Zhong, McDermott  \&
  Surdutovich}{Cholis et~al.}{2022a}]{cholis_2022_6423495}
Cholis I.,  Zhong Y.-M.,  McDermott S.~D.,   Surdutovich J.~P.,  2022a, {The
  Return of the Templates: Revisiting the Galactic Center Excess with
  Multi-Messenger Observations}, \mn@doi{10.5281/zenodo.6423495}, \url
  {https://doi.org/10.5281/zenodo.6423495}

\bibitem[\protect\citeauthoryear{Cholis, Zhong, McDermott  \&
  Surdutovich}{Cholis et~al.}{2022b}]{Cholis2022}
Cholis I.,  Zhong Y.-M.,  McDermott S.~D.,   Surdutovich J.~P.,  2022b, \mn@doi
  [Phys. Rev. D] {10.1103/PhysRevD.105.103023}, 105, 103023

\bibitem[\protect\citeauthoryear{Coleman, Paterson, Gordon, Macias  \&
  Ploeg}{Coleman et~al.}{2020}]{Coleman:2019kax}
Coleman B.,  Paterson D.,  Gordon C.,  Macias O.,   Ploeg H.,  2020, \mn@doi
  [Mon. Not. Roy. Astron. Soc.] {10.1093/mnras/staa1281}, 495, 3350

\bibitem[\protect\citeauthoryear{Daylan, Finkbeiner, Hooper, Linden, Portillo,
  Rodd  \& Slatyer}{Daylan et~al.}{2016}]{Daylan:2014rsa}
Daylan T.,  Finkbeiner D.~P.,  Hooper D.,  Linden T.,  Portillo S. K.~N.,  Rodd
  N.~L.,   Slatyer T.~R.,  2016, \mn@doi [Phys. Dark Univ.]
  {10.1016/j.dark.2015.12.005}, 12, 1

\bibitem[\protect\citeauthoryear{Di~Mauro}{Di~Mauro}{2021}]{DiMauro2021}
Di~Mauro M.,  2021, \mn@doi [Phys. Rev. D] {10.1103/PhysRevD.103.063029}, 103,
  063029

\bibitem[\protect\citeauthoryear{Dinsmore \& Slatyer}{Dinsmore \&
  Slatyer}{2022}]{Dinsmore:2021nip}
Dinsmore J.~T.,  Slatyer T.~R.,  2022, \mn@doi [JCAP]
  {10.1088/1475-7516/2022/06/025}, 06, 025

\bibitem[\protect\citeauthoryear{Feroz, Hobson  \& Bridges}{Feroz
  et~al.}{2009}]{Feroz:2008xx}
Feroz F.,  Hobson M.~P.,   Bridges M.,  2009, \mn@doi [Mon. Not. R. Astron.
  Soc.] {10.1111/j.1365-2966.2009.14548.x}, 398, 1601

\bibitem[\protect\citeauthoryear{Freudenreich}{Freudenreich}{1998}]{Freudenreich:1997bx}
Freudenreich H.~T.,  1998, \mn@doi [Astrophys. J.] {10.1086/305065}, 492, 495

\bibitem[\protect\citeauthoryear{Gautam, Crocker, Ferrario, Ruiter, Ploeg,
  Gordon  \& Macias}{Gautam
  et~al.}{2022}]{gautamMillisecondPulsarsAccretion2022}
Gautam A.,  Crocker R.~M.,  Ferrario L.,  Ruiter A.~J.,  Ploeg H.,  Gordon C.,
   Macias O.,  2022, \mn@doi [Nat. Astron.] {10.1038/s41550-022-01658-3}, 6,
  703

\bibitem[\protect\citeauthoryear{Goodenough \& Hooper}{Goodenough \&
  Hooper}{2009}]{Goodenough:2009gk}
Goodenough L.,  Hooper D.,  2009, \mn@doi [arXiv] {10.48550/arXiv.0910.2998},
  \href {https://ui.adsabs.harvard.edu/abs/2009arXiv0910.2998G} {p.
  arXiv:0910.2998}

\bibitem[\protect\citeauthoryear{Gordon \& Macias}{Gordon \&
  Macias}{2013}]{Gordon:2013vta}
Gordon C.,  Macias O.,  2013, \mn@doi [Phys. Rev. D]
  {10.1103/PhysRevD.88.083521}, 88, 083521

\bibitem[\protect\citeauthoryear{Grand \& White}{Grand \&
  White}{2022}]{Grand:2022olu}
Grand R. J.~J.,  White S. D.~M.,  2022, \mn@doi [Mon. Not. Roy. Astron. Soc.]
  {10.1093/mnrasl/slac011}, 511, L55

\bibitem[\protect\citeauthoryear{Haggard, Heinke, Linden, Hooper  \&
  Linden}{Haggard et~al.}{2017}]{Haggard2017}
Haggard D.,  Heinke C.,  Linden T.,  Hooper D.,   Linden T.,  2017, \mn@doi [J.
  Cosmol. Astropart. Phys.] {10.1088/1475-7516/2017/05/056}, 2017, 0

\bibitem[\protect\citeauthoryear{Hooper \& Goodenough}{Hooper \&
  Goodenough}{2011}]{Hooper:2010mq}
Hooper D.,  Goodenough L.,  2011, \mn@doi [Phys. Lett. B]
  {10.1016/j.physletb.2011.02.029}, 697, 412

\bibitem[\protect\citeauthoryear{Hooper \& Linden}{Hooper \&
  Linden}{2011}]{Hooper:2011ti}
Hooper D.,  Linden T.,  2011, \mn@doi [Phys. Rev. D]
  {10.1103/PhysRevD.84.123005}, 84, 123005

\bibitem[\protect\citeauthoryear{Hooper \& Mohlabeng}{Hooper \&
  Mohlabeng}{2016}]{hooperGammarayLuminosityFunction2016}
Hooper D.,  Mohlabeng G.,  2016, J. Cosmol. Astropart. Phys., 2016, 049

\bibitem[\protect\citeauthoryear{Hooper \& Slatyer}{Hooper \&
  Slatyer}{2013}]{Hooper:2013rwa}
Hooper D.,  Slatyer T.~R.,  2013, \mn@doi [Phys. Dark Univ.]
  {10.1016/j.dark.2013.06.003}, 2, 118

\bibitem[\protect\citeauthoryear{Launhardt, Zylka  \& Mezger}{Launhardt
  et~al.}{2002}]{2002AA...384..112L}
Launhardt R.,  Zylka R.,   Mezger P.~G.,  2002, \mn@doi [Astronomy \&
  Astrophysics] {10.1051/0004-6361:20020017}, \href
  {https://ui.adsabs.harvard.edu/abs/2002A\&A...384..112L} {384, 112}

\bibitem[\protect\citeauthoryear{Lazar et~al.}{Lazar
  et~al.}{2020}]{Lazar:2020pjs}
Lazar A.,  et~al., 2020, \mn@doi [Mon. Not. Roy. Astron. Soc.]
  {10.1093/mnras/staa2101}, 497, 2393

\bibitem[\protect\citeauthoryear{Leane \& Slatyer}{Leane \&
  Slatyer}{2019}]{Leane:2019xiy}
Leane R.~K.,  Slatyer T.~R.,  2019, \mn@doi [Phys. Rev. Lett.]
  {10.1103/PhysRevLett.123.241101}, 123, 241101

\bibitem[\protect\citeauthoryear{Leane \& Slatyer}{Leane \&
  Slatyer}{2020a}]{Leane:2020pfc}
Leane R.~K.,  Slatyer T.~R.,  2020a, \mn@doi [Phys. Rev. D]
  {10.1103/PhysRevD.102.063019}, 102, 063019

\bibitem[\protect\citeauthoryear{Leane \& Slatyer}{Leane \&
  Slatyer}{2020b}]{Leane:2020nmi}
Leane R.~K.,  Slatyer T.~R.,  2020b, \mn@doi [Phys. Rev. Lett.]
  {10.1103/PhysRevLett.125.121105}, 125, 121105

\bibitem[\protect\citeauthoryear{Lee, Lisanti, Safdi, Slatyer  \& Xue}{Lee
  et~al.}{2016}]{Lee:2015fea}
Lee S.~K.,  Lisanti M.,  Safdi B.~R.,  Slatyer T.~R.,   Xue W.,  2016, \mn@doi
  [Phys. Rev. Lett.] {10.1103/PhysRevLett.116.051103}, 116, 051103

\bibitem[\protect\citeauthoryear{Linden, Rodd, Safdi  \& Slatyer}{Linden
  et~al.}{2016}]{Linden:2016rcf}
Linden T.,  Rodd N.~L.,  Safdi B.~R.,   Slatyer T.~R.,  2016, \mn@doi [Phys.
  Rev. D] {10.1103/PhysRevD.94.103013}, 94, 103013

\bibitem[\protect\citeauthoryear{List, Rodd, Lewis  \& Bhat}{List
  et~al.}{2020}]{List:2020mzd}
List F.,  Rodd N.~L.,  Lewis G.~F.,   Bhat I.,  2020, \mn@doi [Phys. Rev.
  Lett.] {10.1103/PhysRevLett.125.241102}, 125, 241102

\bibitem[\protect\citeauthoryear{List, Rodd  \& Lewis}{List
  et~al.}{2021}]{List:2021aer}
List F.,  Rodd N.~L.,   Lewis G.~F.,  2021, \mn@doi [Phys. Rev. D]
  {10.1103/PhysRevD.104.123022}, 104, 123022

\bibitem[\protect\citeauthoryear{Macias, Gordon, Crocker, Coleman, Paterson,
  Horiuchi  \& Pohl}{Macias et~al.}{2018}]{Macias:2016nev}
Macias O.,  Gordon C.,  Crocker R.~M.,  Coleman B.,  Paterson D.,  Horiuchi S.,
    Pohl M.,  2018, \mn@doi [Nat. Astron.] {10.1038/s41550-018-0414-3}, 2, 387

\bibitem[\protect\citeauthoryear{Macias, Horiuchi, Kaplinghat, Gordon, Crocker
  \& Nataf}{Macias et~al.}{2019}]{Macias:2019omb}
Macias O.,  Horiuchi S.,  Kaplinghat M.,  Gordon C.,  Crocker R.~M.,   Nataf
  D.~M.,  2019, \mn@doi [J. Cosmol. Astropart. Phys.]
  {10.1088/1475-7516/2019/09/042}, 09, 042

\bibitem[\protect\citeauthoryear{Macias, van Leijen, Song, Ando, Horiuchi  \&
  Crocker}{Macias et~al.}{2021}]{Macias:2021boz}
Macias O.,  van Leijen H.,  Song D.,  Ando S.,  Horiuchi S.,   Crocker R.~M.,
  2021, \mn@doi [Mon. Not. Roy. Astron. Soc.] {10.1093/mnras/stab1450}, 506,
  1741

\bibitem[\protect\citeauthoryear{Malyshev}{Malyshev}{2024}]{Malyshev24}
Malyshev D.~V.,  2024, \mn@doi [arXiv]
  {https://doi.org/10.48550/arXiv.2401.04565}, p. arXiv:2401.04565

\bibitem[\protect\citeauthoryear{Manconi, Calore  \& Donato}{Manconi
  et~al.}{2024}]{Manconi:2024DMlimits}
Manconi S.,  Calore F.,   Donato F.,  2024, in preparation

\bibitem[\protect\citeauthoryear{McCann}{McCann}{2015}]{McCann:2014dea}
McCann A.,  2015, \mn@doi [Astrophys. J.] {10.1088/0004-637X/804/2/86}, 804, 86

\bibitem[\protect\citeauthoryear{{McDermott}, {Zhong}  \& {Cholis}}{{McDermott}
  et~al.}{2023}]{McDermott2023}
{McDermott} S.~D.,  {Zhong} Y.-M.,   {Cholis} I.,  2023, \mn@doi [Mon. Not. R.
  Astron. Soc.] {10.1093/mnrasl/slad035}, \href
  {https://ui.adsabs.harvard.edu/abs/2023MNRAS.522L..21M} {522, L21}

\bibitem[\protect\citeauthoryear{Mishra-Sharma \& Cranmer}{Mishra-Sharma \&
  Cranmer}{2022}]{Mishra-Sharma:2021oxe}
Mishra-Sharma S.,  Cranmer K.,  2022, \mn@doi [Phys. Rev. D]
  {10.1103/PhysRevD.105.063017}, 105, 063017

\bibitem[\protect\citeauthoryear{Nishiyama et~al.}{Nishiyama
  et~al.}{2013}]{Nishiyama2015}
Nishiyama S.,  et~al., 2013, \mn@doi [Astrophys. J. Lett.]
  {10.1088/2041-8205/769/2/L28}, 769, L28

\bibitem[\protect\citeauthoryear{Ploeg, Gordon, Crocker  \& Macias}{Ploeg
  et~al.}{2020}]{ploegComparingGalacticBulge2020}
Ploeg H.,  Gordon C.,  Crocker R.,   Macias O.,  2020, \mn@doi [JCAP]
  {10.1088/1475-7516/2020/12/035}, 12, 035

\bibitem[\protect\citeauthoryear{Pohl, Macias, Coleman  \& Gordon}{Pohl
  et~al.}{2022}]{2022ApJ...929..136P}
Pohl M.,  Macias O.,  Coleman P.,   Gordon C.,  2022, \mn@doi [Astrophys. J.]
  {10.3847/1538-4357/ac6032}, 929, 136

\bibitem[\protect\citeauthoryear{Porter, Johannesson  \& Moskalenko}{Porter
  et~al.}{2017}]{Porter:2017vaa}
Porter T.~A.,  Johannesson G.,   Moskalenko I.~V.,  2017, \mn@doi [Astrophys.
  J.] {10.3847/1538-4357/aa844d}, 846, 67

\bibitem[\protect\citeauthoryear{Song, Macias  \& Horiuchi}{Song
  et~al.}{2019}]{Song:2019nrx}
Song D.,  Macias O.,   Horiuchi S.,  2019, \mn@doi [Phys. Rev. D]
  {10.1103/PhysRevD.99.123020}, 99, 123020

\bibitem[\protect\citeauthoryear{Storm, Weniger  \& Calore}{Storm
  et~al.}{2017}]{Storm:2017arh}
Storm E.,  Weniger C.,   Calore F.,  2017, \mn@doi [J. Cosmol. Astropart.
  Phys.] {10.1088/1475-7516/2017/08/022}, 1708, 022

\bibitem[\protect\citeauthoryear{Strong \& Moskalenko}{Strong \&
  Moskalenko}{1998}]{Strong:1998pw}
Strong A.~W.,  Moskalenko I.~V.,  1998, \mn@doi [Astrophys. J.]
  {10.1086/306470}, 509, 212

\bibitem[\protect\citeauthoryear{Trotta}{Trotta}{2008}]{Trotta:2008qt}
Trotta R.,  2008, \mn@doi [Contemp. Phys.] {10.1080/00107510802066753}, 49, 71

\bibitem[\protect\citeauthoryear{{Vitale} \& {Morselli}}{{Vitale} \&
  {Morselli}}{2009}]{Vitale2009}
{Vitale} V.,  {Morselli} A.,  2009, \mn@doi [arXiv] {10.48550/arXiv.0912.3828},
  \href {https://ui.adsabs.harvard.edu/abs/2009arXiv0912.3828V} {p.
  arXiv:0912.3828}

\bibitem[\protect\citeauthoryear{Weiland, Arendt, Berriman  \& {...}}{Weiland
  et~al.}{1994}]{WeilandCOBEDiffuseInfrared1994}
Weiland J.~L.,  Arendt R.~G.,  Berriman G.~B.,   {...} 1994, \mn@doi
  [Astrophys. J.] {10.1086/187315}, \href
  {https://ui.adsabs.harvard.edu/abs/1994ApJ...425L..81W} {425, L81}

\bibitem[\protect\citeauthoryear{Wolleben}{Wolleben}{2007}]{2007ApJ...664..349W}
Wolleben M.,  2007, \mn@doi [Astrophys. J.] {10.1086/518711}, \href
  {https://ui.adsabs.harvard.edu/abs/2007ApJ...664..349W} {664, 349}

\bibitem[\protect\citeauthoryear{Zhong \& Cholis}{Zhong \&
  Cholis}{2024}]{ZhongCholis24}
Zhong Y.-M.,  Cholis I.,  2024, \mn@doi [arXiv]
  {https://doi.org/10.48550/arXiv.2401.02481}, p. arXiv:2401.02481

\bibitem[\protect\citeauthoryear{Zhou, Liang, Huang, Li, Fan, Feng  \&
  Chang}{Zhou et~al.}{2015}]{Zhou:2014lva}
Zhou B.,  Liang Y.-F.,  Huang X.,  Li X.,  Fan Y.-Z.,  Feng L.,   Chang J.,
  2015, \mn@doi [Phys. Rev. D] {10.1103/PhysRevD.91.123010}, 91, 123010

\makeatother
\end{thebibliography}

\appendix

\bsp	
\label{lastpage}
\end{document}